\documentclass[11pt]{article}
\usepackage{color}
\usepackage{amssymb}
\usepackage{graphicx}
\usepackage{epsfig,rotating}
\usepackage{exscale}
\usepackage{amsmath}
\usepackage{cite}
\usepackage{latexsym}
\usepackage{multirow}
\usepackage{graphics}
\usepackage{cite}
\usepackage[english]{babel}
\newcommand{\bm}{\begin{minipage}}
\renewcommand{\em}{\end{minipage}}
\newcommand{\bi}{\begin{itemize}}
\newcommand{\ei}{\end{itemize}}
\newcommand{\be}{\begin{eqnarray}}
\newcommand{\en}{\end{eqnarray}}
\newcommand{\ba}{\begin{array}}
\newcommand{\ea}{\end{array}}
\newcommand{\bc}{\begin{center}}
\newcommand{\ec}{\end{center}}
\newcommand{\bfi}{\begin{figure}}
\newcommand{\efi}{\end{figure}}
\newcommand{\im}{\mbox{Im}}
\newcommand{\re}{\mbox{Re}}
\newcommand{\bfr}{\begin{flushright}}
\newcommand{\efr}{\end{flushright}}

\newcommand{\no}{\nonumber}

\newcommand{\ov}[1]{\overline{#1}}

\newcommand{\bbr}{{\it B{\scriptsize A}B{\scriptsize AR}}}

\newcommand{\ds}{\displaystyle}
\def\ees+s+        {\ensuremath{e^+ e^- \!\rightarrow \Sigma^+\overline{\Sigma^+}}}

\def\ee        {\ensuremath{e^+e^-}}

\def\gev       {\ensuremath{{\rm GeV}}}

\def\mr        {\ensuremath{M_{\rho}}}

\def\gr        {\ensuremath{\Gamma^{\rho}}}
\def\mpi        {\ensuremath{M_{\pi}}}
\def\NN        {\ensuremath{N\overline{N}}}
\def\fis        {\ensuremath{F^{\rm is}}}
\def\fiv        {\ensuremath{F^{\rm iv}}}

\begin{document}

\title{Time-like and space-like
electromagnetic form factors of nucleons,
a unified description}
\author{Earle L. Lomon\\
\small\it  Center for Theoretical Physics and Laboratory for Nuclear Science\\
\small\it  and Department of Physics, Massachusetts Institute of Technology,\\
\small\it  Cambridge, Massachusetts 02139
\vspace{5mm}\\
Simone Pacetti\\
\small\it Department of Physics, University of Perugia\\ 
\small\it and I.N.F.N. Perugia, Italy}
\date{}
\maketitle

\begin{abstract}
The extended Lomon-Gari-Kr\"umpelmann model of nucleon electromagnetic form factors, 
which embodies $\rho$, $\rho'$, $\omega$, $\omega'$ and $\phi$ vector meson 
contributions and the perturbative QCD high momentum transfer behavior has been
extended to the time-like region. Breit-Wigner formulae with momentum-dependent
widths have been considered for broad resonances in order to have a parametrization 
for the electromagnetic form factors that fulfills, in the time-like region, constraints 
from causality, analyticity, and unitarity.

This analytic extension of the Lomon-Gari-Kr\"umpelmann model has been used to
perform a unified fit to all the nucleon electromagnetic form factor data, in the
space-like and time-like region (where form factor values are extracted from
$\ee\leftrightarrow\NN$ cross sections data).

The knowledge of the complete analytic structure of form factors enables
predictions at extended momentum transfer, and also of time-like observables such as the ratio between electric and magnetic
form factors and their relative phase.
\end{abstract}

\section{Introduction}
\label{sec:intro}
Nucleon electromagnetic form factors (EMFF's) describe modifications of 
the pointlike photon-nucleon vertex due to the structure of nucleons. 
Because the virtual photon interacts with single elementary charges, 
the quarks, it is a powerful probe for the internal structure of composite 
particles.  Moreover, as the electromagnetic interaction is 
precisely calculable in QED, the dynamical content of each vertex can 
be compared with the data.\\
The study of EMFF's is an essential step towards a deep understanding 
of the low-energy QCD dynamics. Nevertheless, even in case of nucleons,
the available data are still incomplete. \\
The experimental situation is twofold: 
\bi
\item in the space-like region many data sets are available for elastic 
electron scattering from nucleons ($N$), both protons ($p$) and 
neutrons ($n$).  Recently,  the development of new polarization techniques 
(see e.g. Ref.~\cite{perdrisat}) provides an important improvement to the accuracy, 
giving a better capability of disentangling electric and magnetic EMFF's than 
the unpolarized differential cross sections alone.
\item In the time-like region there are few measurements, mainly of the total 
cross section (in a restricted angular range) of $\ee\leftrightarrow\NN$,  one set 
for neutrons and nine sets for protons, one of which includes a produced photon. 
Only two attempts, with incompatible results, have been made to separate the 
electric and magnetic EMFF's in the time-like region.
\ei
Many models and 
interpretations for the nucleon EMFF's have been proposed. 
Such a wide variety of descriptions reflects the difficulty 
of connecting the phenomenological properties of nucleons, parametrized
by the EMFF's, to the underlying theory which is QCD in the
non-perturbative (low-energy) regime.\\
The analyticity requirement, which connects descriptions 
 in both space ($q^2<0$) and time-like ($q^2>0$) regions, 
drastically reduces the range of models to be considered. 
In particular, the more successful ones in the space-like region are the 
Vector-Meson-Dominance (VMD) based models~\cite{IJL,gari-krumpe}
(see, e.g., Ref.~\cite{vmd} for a review on VMD models) 
that, in addition, because of their analytic form, have the
property of being easily extendable to the whole $q^2$-domain: space-like, time-like 
and asymptotic regions.\\
In this paper we propose an analytic continuation to the time-like region
of the last version of the Lomon model for the space-like nucleon EMFF's~\cite{lomon-list}.
This model has been developed by improving the original idea, due to Iachello, 
Jackson and Land\'e~\cite{IJL} and further developed by
Gari and Kr\"umpelmann~\cite{gari-krumpe}, who gave a description of
nucleon EMFF's which incorporates: VMD at low momentum
transfer and asymptotic freedom in the perturbative QCD (pQCD) regime.
As we will see in Sec.~\ref{sec:model}, in this model
EMFF's are described by two kinds of functions: vector meson propagators,
dominant at low-$q^2$ and hadronic form factors (FF's) at high-$q^2$. The 
analytic extension of the model only modifies the propagator part and
consists in defining more accurate expressions for propagators 
that account for finite-width effects and give the expected resonance
singularities in the $q^2$-complex plane.

\section{Nucleon electromagnetic form factors}
\label{sec:Nemff}
The elastic scattering of an electron by a nucleon $e^- N\to e^- N$
is represented, in Born approximation, by the diagram 
of Fig.~\ref{fig:ee-nn} in the vertical direction.
In this kinematic region the 4-momentum of the virtual photon is
space-like: $q^2=-2\omega_1\omega_2(1-\cos\theta_e)\le 0$, 
$\omega_{1(2)}$ is the energy of the incoming (outgoing) electron 
and $\theta_e$ is the scattering angle.
%
%  Feynman diagram ee-NN
%
\bfi[h!]
\centering
\bm{93mm}\centering
\epsfig{file=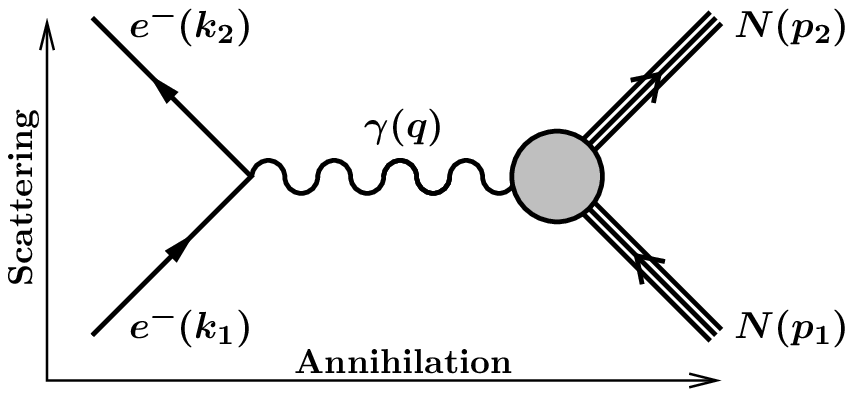,width=90mm}
\caption{One-photon exchange Feynman diagram for 
scattering $e^- N\to e^- N$ and annihilation
$\ee\to\NN$.}
\label{fig:ee-nn}\em
\efi\\
The annihilation $\ee\to\NN$ or $\NN\to\ee$ is represented
by the same diagram of Fig.~\ref{fig:ee-nn} but in the horizontal
direction, in this case the 4-momentum $q$ is time-like: 
$q^2=(2\omega)^2\ge 0$, where $\omega\equiv \omega_1=
\omega_2$ is the common value of the lepton energy in the \ee\ 
center of mass frame.\\
The Feynman amplitude for the elastic scattering is
\be
\mathcal{M}=\frac{1}{q^2}\big[e\,\ov{u}(k_2)\gamma^\mu u(k_1)\big]
\big[e\,\ov{U}(p_2)\Gamma_\mu(p_1,p_2) U(p_1)\big]\,,
\no\label{eq:amp}
\en
where the 4-momenta follow the labelling of Fig.~\ref{fig:ee-nn},
$u$ and $U$ are the electron and nucleon spinors, and $\Gamma^\mu$
is a non-constant matrix which describes the nucleon vertex. Using
gauge and Lorentz invariance the most general form of such a matrix
is~\cite{foldy}
\be
\Gamma^\mu=\gamma^\mu F_1^N(q^2)+\frac{i\sigma^{\mu\nu}q_\nu}{2M_N}F_2^N(q^2)\,,
\label{eq:Gamma}
\en
where $M_N$ is the nucleon mass ($N=n$, $p$), $F_{1}^N(q^2)$ 
and $F_{2}^N(q^2)$ are the so-called Dirac and Pauli EMFF's, they are 
Lorentz scalar functions of $q^2$ and describe the non-helicity-flip 
and the helicity-flip part of the hadronic current respectively. 
Normalizations at $q^2=0$ follow from total charge and
static magnetic moment conservation and are
\be
F_1^N(0)=Q_N\,,\hspace{5mm}
F_2^N(0)=\kappa_N\,,
\no\label{eq:norm}
\en 
where $Q_N$ is the electric charge (in units of $e$) and $\kappa_N$ 
the anomalous magnetic moment (in units of the Bohr magneton $\mu_B$)
of the nucleon $N$.\\
In the Breit frame, i.e. when the transferred 4-momentum $q$ is 
purely space-like, $q=(0,\vec{q})$, the hadronic current takes the standard
form of an electromagnetic 4-current, where the time and
the space component are Fourier transformations of a charge 
and a current density respectively:
\be
\left\{\begin{array}{l}
\rho_q=J^0=\displaystyle e\left[F_1^N+\frac{q^2}{4M^2} F_2^N\right]\\
\\
\vec{J}_q=e\,\ov{U}(p_2)\vec{\gamma}U(p_1)\,[F_1^N+F_2^N]\\
\end{array}\right.\,.
\no\label{eq:sachs0}
\en  
We can define another pair of EMFF's through the
combinations
\be
\left\{\begin{array}{l}
\displaystyle G_E^N=F_1^N+\frac{q^2}{4M_N^2} F_2^N\\
\\
\displaystyle G_M^N=F_1^N+ F_2^N\, \end{array}\right.
\label{eq:sachs}
\en 
these are the Sachs electric and magnetic EMFF's~\cite{Hand:1963zz},
that, in the Breit frame, correspond to the Fourier transformations of the 
charge and magnetic moment spatial distributions of the nucleon. The normalizations,
which reflect this interpretation, are 
\be
G_E^N(0)=Q_N\,,\hspace{5mm}
G_M^N(0)=\mu_N\,,
\no\label{eq:norm2}
\en 
where $\mu_N=Q_N+\kappa_N$ is the nucleon magnetic moment.
Moreover, Sachs EMFF's are equal to each other at the time-like production 
threshold  $q^2=4M_N^2$, i.e.:
\be
G_E^N(4M_N^2)=G_M^N(4M_N^2)\,.
\no\label{eq:th-indentity}
\en
Finally, we can consider the isospin decomposition for the Dirac and Pauli
EMFF's
\be
\fis_i=\frac{1}{2}(F_i^p+F_i^n)\,,
\hspace{20mm}
\fiv_i=\frac{1}{2}(F_i^p-F_i^n)\,,
\hspace{20mm}i=1,\,2\,,
\label{eq:iso-comp}
\en 
\fis\ and \fiv\ are the isoscalar and isovector components.

\section{The Model}
\label{sec:model}
The model presented here is based on simpler versions designed for the 
space-like EMFF's of Iachello, Jackson and Land\'e~\cite{IJL} and of Gari 
and Kr\"umpelmann idea~\cite{gari-krumpe},
which describes nucleon EMFF's by means of a mixture 
of VMD, for the electromagnetic low-energy part, and strong 
vertex FF's for the asymptotic behavior of super-convergent or pQCD.  
The Lomon version~\cite{lomon-list}, which fits well all the 
space-like data now available included two more well identified 
vector mesons and an analytic correction to the form of the  
$\rho$ meson propagator suitable for describing the effect of 
its decay width in the space-like region  fitted to a dispersive 
analysis by Mergell, Meissner, and Drechsel~\cite{meissner}.\\
This model describes the isospin components, eq.~(\ref{eq:iso-comp}),
in order to separate different species of vector meson contributions. For the isovector part
the Lomon model used the $\rho$ and $\rho(1450)$ or $\rho'$ contribution, while for the isoscalar the
$\omega$, $\omega(1420)$ or $\omega'$ and $\phi$ were considered. In detail these are the 
expressions:
\be
\begin{array}{rcl}
\fiv_1(q^2)&\!\!\!\!=\!\!\!\!&\ds \big[BW_{\rm MMD}^{1,\rho}(q^2)\,+BW_0^{\rho'}(q^2)\,\big]F_1^\rho(q^2)+
\big[1-BW_{\rm MMD}^{1,\rho}(0)-BW_0^{\rho'}(0)\big]F_1^D(q^2)\\
&\!\!\!\!\!\!\!\!&\\
\fiv_2(q^2)&\!\!\!\!=\!\!\!\!&\ds\kappa_\rho\, \big[BW_{\rm MMD}^{2,\rho}(q^2)\,+
BW_0^{\rho'}(q^2)\,\big]F_2^\rho(q^2)+
\big[1-BW_{\rm MMD}^{2,\rho}(0)-BW_0^{\rho'}(0)\big]F_2^D(q^2)\\
&\!\!\!\!\!\!\!\!&\\
\fis_1(q^2)&\!\!\!\!=\!\!\!\!&\ds \big[BW_0^\omega(q^2) +BW_0^{\omega'}(q^2)\,\big]F_1^\omega(q^2)+
BW_0^\phi(q^2)\,F_1^\phi(q^2)+\\
&\!\!\!\!\!\!\!\!&\big[1-BW_0^\omega(0)\,-BW_0^{\omega'}(0)\big]F_1^D(q^2)\\
&\!\!\!\!\!\!\!\!&\\
\fis_2(q^2)&\!\!\!\!=\!\!\!\!&\ds \big[\kappa_\omega\,BW_0^\omega(q^2)\,+\kappa_{\omega'}\,
BW_0^{\omega'}(q^2)\,\big]F_2^\omega(q^2)+
\kappa_\phi\,BW_0^\phi(q^2)\,F_2^\phi(q^2)+\\
&\!\!\!\!\!\!\!\!&\big[\kappa_{\rm s}-\kappa_\omega\,BW_0^\omega(0)-
\kappa_{\omega'}\,BW_0^{\omega'}(0)\big]F_2^D(q^2)\,,\\
\label{eq:gk-ff}
\end{array}
\en
where:
\bi
\item $BW_0^\alpha(q^2)$ is the propagator of the intermediate vector meson 
$\alpha$ in pole approximation
\be
BW_0^\alpha(q^2)=\frac{g_\alpha}{f_\alpha}\,\frac{M_\alpha^2}{M_\alpha^2-q^2}\,,
\hspace{15mm}
\alpha=\rho',\omega,\omega',\phi\,,
\label{eq:bw0}
\en
$M_\alpha^2\, g_\alpha/f_\alpha$ are the couplings to the virtual photon and the nucleons;
\item $BW_{\rm MMD}^{i,\rho}(q^2)$ are dispersion-integral analytic approximations 
for the $\rho$  meson contribution in the space-like region~\cite{meissner} 
\be
BW_{\rm MMD}^{1,\rho}(q^2)&\!\!\!\!=\!\!\!\!&
\frac{1.0317 + 0.0875\,(1 -q^2/0.3176)^{-2}}{2\,(1 -q^2 /0.5496)}\,,
\no\\
BW_{\rm MMD}^{2,\rho}(q^2)&\!\!\!\!=\!\!\!\!&
\frac{5.7824 + 0.3907\,(1 -q^2/0.1422)^{-1}}{2\,\kappa_\rho(1 -q^2 /0.5362)}\,;
\no\en
\item 
the last term in each expression of eq.~(\ref{eq:gk-ff}) dominates the 
asymptotic QCD behavior and also normalizes the EMFF's at $q^2=0$ to the 
charges and anomalous magnetic moments of the nucleons;
\item the functions $F_i^\alpha(q^2)$, $\alpha=\rho,\omega,\phi$ and $i=1,2$, are 
meson-nucleon FF's which describe the vertices $\alpha NN$, where a virtual 
vector meson $\alpha$ couples with two on-shell nucleons.  
Noting that the same meson-nucleon FF's are used for $\rho'$ and $\omega'$ as for 
$\rho$ and $\omega$, we have
\be
\begin{array}{rcl}
F_i^{\rho,\omega}(q^2)&=&\ds f_i(q^2)\equiv
\frac{\Lambda_1^2}{\Lambda_1^2- q^2}
\left(\frac{\Lambda_2^2}{\Lambda_2^2- q^2}\right)^i,\hspace{10mm}i=1,2\,,\\
&&\\
F_1^\phi(q^2)&=&\ds f_1(q^2)\left(\frac{ q^2}{ q^2-\Lambda_1^2}\right)^{3/2}\,, \\
&&\\
F_2^\phi(q^2)&=&\ds f_2(q^2)\left(\frac{\Lambda_1^2}{\mu_\phi^2}\,
\frac{ q^2-\mu_\phi^2}{ q^2-\Lambda_1^2}\right)^{3/2}\,, \\
\end{array}
\label{eq:fi}
\en
where $\Lambda_1$ and $\Lambda_2$ are free parameters that represent
cut-offs for the general high energy behavior and the helicity-flip
respectively, and
\be
\tilde q^2=q^2\frac{\ln\big[(\Lambda_D^2-q^2)/\Lambda_{\rm QCD}^2\big]}
{\ln\big(\Lambda_D^2/\Lambda_{\rm QCD}^2\big)}\,,
\label{eq:q2tilde}
\en
where $\Lambda_{D}$ is another free cut-off
which controls the asymptotic behavior of the quark-nucleon vertex,
the extra factor in $F_i^\phi(q^2)$ imposes the Zweig rule;
\item the functions $F_i^D(q^2)$ can be interpreted as quark-nucleon FF's that parametrize 
the direct coupling of the virtual photon to the valence quarks of the nucleons, 
\be
F_i^D(q^2)=\frac{\Lambda_D^2}{\Lambda_D^2-\tilde q^2}
\left(\frac{\Lambda_2^2}{\Lambda_2^2-\tilde q^2}\right)^i,\hspace{10mm}i=1,2\,,
\label{eq:f12D}
%\no
\en
$\tilde q^2$ is defined as in eq.~(\ref{eq:q2tilde}); %;
\item
finally, $\kappa_\alpha$ is the ratio of tensor to vector coupling
at $q^2=0$ in the $\alpha NN$ matrix element, while the isospin 
anomalous magnetic moments are
\be
\kappa_{\rm s}=\kappa_p+\kappa_n\,,\hspace{20mm}
\kappa_{\rm v}=\kappa_p-\kappa_n\,.
\no
\en  
\ei
The space-like asymptotic behavior ($q^2\to-\infty$) for the Dirac and Pauli EMFF's
of eq.~(\ref{eq:gk-ff}) is driven by the $F_{1,2}^D(q^2)$ contribution, given in 
eq.~(\ref{eq:f12D}).
In particular we get
\be
\begin{array}{rcl}
F_1^{\rm iv,is}(q^2)&\ds\mathop{\sim}_{q^2\to-\infty}&\ds
\frac{1}{\Big[q^2\ln\big(-q^2/\Lambda_{\rm QCD}^2\big)\Big]^2}\\
&&\\
F_2^{\rm iv,is}(q^2)&\ds\mathop{\sim}_{q^2\to-\infty}&\ds
\frac{F_1^{\rm iv,is}(q^2)}{-q^2\ln\big(-q^2/\Lambda_{\rm QCD}^2\big)}\,,\\
\end{array}
\no
\en
as required by the pQCD~\cite{pqcd}.
\\
% 
%\ei
%
In principle this model can be extended also to the time-like 
region, positive $q^2$, to describe data on cross sections for
the annihilation processes: $\ee\leftrightarrow\NN$. However,
a simple analytic continuation of the expressions given in
eq.~(\ref{eq:gk-ff}) involves important issues mainly concerning 
the analytic structure of the vector meson components of the EMFF's 
that, in the time-like region, are complex functions 
of $q^2$. 
%
% added 16th may
%
The hadronic FF's of eqs.~(\ref{eq:fi}) and~(\ref{eq:f12D}) may also 
have real poles as a function of $q^2$.  In fact as defined above $F_i^{\phi}$ 
has a real pole at $q^2=\Lambda_i^2$.  
In the other denominators of eqs.~(\ref{eq:fi}) and~(\ref{eq:f12D}), as in 
$F_i^{\rho,\omega}$ and $F_1^D$, $q^2$ is replaced by $\tilde q^2$.  
The latter as a function of $q^2$ has a maximum in its real 
range $0<q^2<\Lambda_D^2$, which, for reasonable values of $\Lambda_D$ and 
$\Lambda_{\rm QCD}$, may be smaller than $\Lambda_1^2$, $\Lambda_2^2$ and 
$\Lambda_D^2$.  Therefore all the hadronic FF's real poles may be avoided by 
also replacing $q^2$ by $\tilde q^2$ in the factors of $F_i^{\phi}$. 
This does not effect the asymptotic behavior required by the Zweig rule and 
will be adopted in the model used here. The results in Sec.~\ref{sec:results} 
show that with this modification real poles can be avoided in every case examined, 
although in half the cases mild constraints on $\Lambda_1$ or $\Lambda_{\rm QCD}$ 
are needed which affect the quality of the fit negligibly.
\\
A detailed treatment of the possibility of extending the model
from the space-like to the time-like region, will be given in 
Sec.~\ref{sec:extension}.
\section{Analyticity of Breit-Wigner formulae}
\label{sec:analytic}
The standard relativistic Breit-Wigner (BW) formula for an unstable
particle of mass $M$ and energy independent width $\Gamma$ is
\be
BW(s)=\frac{1}{M^2-s-i\,\Gamma\, M}\,,\no
\en
it has a very simple analytic structure, only one complex pole and
no discontinuity cut in its domain. Once this formula is improved to include
energy dependent widths one immediately face problems 
concerning the analyticity. \\
We consider explicitly the case of the $\rho$ resonance in
its dominant decay channel $\pi^+\pi^-$. A realistic way to formulate an energy 
dependent width is to extend the $\rho$ mass off-shell, making the 
substitution $\mr^2= s$,  in the first order decay rate
\be
\Gamma(\rho\to\pi^+\pi^-)=\frac{|g^\rho_{\pi\pi}|^2}{48\pi}\,
\frac{\mr^2-4\mpi^2}{\mr^2}\,,
\label{eq:gammarho}
\en
where $g^\rho_{\pi\pi}$ is the coupling constant and, \mr\ and \mpi\
are the $\rho$ and pion mass respectively. 
Such a decay rate has been obtained by considering, for the vertex 
$\rho\pi^+\pi^-$, the pointlike amplitude
\be
\mathcal{M}=g^\rho_{\pi\pi}\,\epsilon_\mu \,
(p_+-p_-)^\mu\,,\no
\en
where $\epsilon_\mu$ is the polarization vector of the vector meson
$\rho$, and $p_{\pm}$ the 4-momentum of $\pi^\pm$.\\ 
Finally, assuming the $\pi^+\pi^-$ as the only decay channel and
using eq.~(\ref{eq:gammarho}) for the corresponding rate, the energy 
dependent width can be defined as
\be
\gr_s(s)=\gr_0\,\frac{\mr^2}{s}\left(
\frac{s-s_0}{\mr^2-s_0}\right)^{\frac{3}{2}}
\equiv \frac{\gamma_\rho}{\mr}\, \frac{\left(s-s_0\right)^{\frac{3}{2}}}{s}\,,
\hspace{10mm}
\gamma_\rho\equiv
\frac{\gr_0 \mr^3}{(\mr^2-s_0)^\frac{3}{2}}\,,
\label{eq:width}
\en
where the subscript ``$s$'' indicates the factor $1/s$ appearing in the
width definition, $\gr_0$ is the total width of the $\rho$, and $s_0=4M_\pi^2$. 
It follows that the BW formula becomes
\be
BW_s(s)=\frac{s}{s(\mr^2-s)-i\gamma_\rho\,\left(s-s_0\right)^\frac{3}{2}}\,.
\no\label{eq:bw}
\en
In this form the BW has the ``required''~\cite{bw} discontinuity cut
$(s_0,\infty)$ and maintains a complex pole $s_p$
\be
s_p= \tilde \mr^2+ i\tilde\gr_0 \,\tilde \mr\simeq 
\mr^2+ i \gr_0 \, \mr=s^0_p\,.\no
\en 
Due to the more complex analytic structure the new pole position $s_p$
turns out to be slightly shifted with respect to the original
position $s_p^0$. Moreover, these are not the only complications introduced by 
using $\gr(s)$ instead of $\gr_0$, the power 3/2 in the denominator
and the factor $1/s$, see eq.~(\ref{eq:width}), generate also additional
physical poles which, in agreement with dispersion relations, must be subtracted, as discussed below. 
\subsection{Regularization of Breit-Wigner formulae}
\label{subsec:regu}
We consider the general case where there is a number $N$ of  
poles lying in the physical Riemann sheet. We may rewrite
the BW by separating the singular and regular behaviors as
% IN THE 2 FOLLOWING EQUATIONS IT MAY BE BETTER TO REPLACE THE
% SUBSCRIPT INDEX i WITH j TO AVOID CONFUSION WITH THE
% IMAGINARY i. DONE!
\be
BW(s)=\frac{P_N(s)}{\prod_{j=1}^N(s-z_j)
\big[M^2-s-i \gamma (s-s_0)^\beta\big]}\,,
\no
\en 
where $P_N(s)$ is a suitable $N$ degree polynomial, 
$\beta$ is a non-integer real number which defines the 
discontinuity cut (in the previous case we had $\beta=3/2$),
$\gamma= M\,\Gamma^0/(M^2-s_0)^\beta$, and the
$z_j$ are the real axis (physical) poles.
To avoid divergences in our formulae, we may define
a simple regularization procedure consisting in subtracting these
poles. In other words we add counterparts that at $z=z_j$ behave 
as the opposite of the $i$-th pole. In more detail, we may
define a regularized BW as
% IN THIS FORMULA I THINK THAT THE INDICES k and i[j] IN THE
% PRODUCT IN DENOMINATOR NEED TO BE INTERCHANGED
% (IT HAS ALREADY BEEN SUMMED OVER k). ALMOST DONE!
\be
\widetilde{BW}(s)=BW(s)-\sum_{k=1}^N
\frac{P_N(z_k)}{\prod_{j=1,j\not=k}^N(z_k-z_j)
\big[{M}^2-z_k-i\gamma(z_k-s_0)^\beta\big]}
\,\frac{1}{s-z_k}\,.
\label{eq:regu0}
\en
In the Appendix~A we show how dispersion relations (DR's)
offer a powerful tool to implement this procedure without the
need to know where the poles are located. 
However in this paper we show that an analytic expression 
also contains the information.
%
%
%
% WOULD THE FOLLOWING SUBSECTION BE MORE SUITABLE AS AN
% APPENDIX? DONE!
%
%
%
%%%%%%%%%%%%%%%%%%%%%%%%%%%%%%%%%%%%%%%%%%%%%%%%%%%%%%%%%%%%%%%%%%%%%%%
% 
\subsection{Two cases for $\Gamma(s)$}
\label{subsec:bws}
In our model for nucleon EMFFs, widths are used only for the
broader resonances: $\rho(770)$, $\rho(1450)$ and $\omega(1420)$~\cite{pdg}.
We consider explicitly two expressions for $\Gamma(s)$ which entail different
analytic structures for the BW formulae. Besides the form we discussed in 
Sec.~\ref{sec:analytic}, eq.~(\ref{eq:width}), we consider also a simpler 
expression (closer to the non-relativistic form), hence
for a generic broad resonance we have
\be
\begin{array}{lrl}
\displaystyle\Gamma_{s}(s)=\Gamma_0\,\frac{M^2}{s}\left(
\frac{s-\tilde{s}_0}{M^2-\tilde{s}_0}\right)^{\frac{3}{2}}
\equiv \frac{\gamma_s}{M}\, \frac{\left(s-\tilde{s}_0\right)^{\frac{3}{2}}}{s}\,,
&& \mbox{with: }\displaystyle\gamma_s=\frac{\Gamma_0 M^3}{\left(M^2-\tilde{s}_0\right)^{\frac{3}{2}}}\\
&\hspace{5mm}&\\
\displaystyle\Gamma_{1}(s)=\Gamma_0\left(
\frac{s-\tilde{s}_0}{M^2-\tilde{s}_0}\right)^{\frac{3}{2}}
\equiv \frac{\gamma_1}{M}\, \left(s-\tilde{s}_0\right)^{\frac{3}{2}}\,,
&& \mbox{with: }\displaystyle\gamma_1=
\frac{\Gamma_0 M}{\left(M^2-\tilde{s}_0\right)^{\frac{3}{2}}}\,.\\
\end{array}
\label{eq:widths}
\en
In both cases we assume that such a resonance decays predominantly
into a two-body channel whose mass squared equals $\tilde{s}_0$.
The subscript ``1'' in the second expression 
of eq.~(\ref{eq:widths}) indicates that there is no extra factor $1/s$ in
the definition of the energy-dependent width.\\
As already discussed, the BW formulae acquire a more complex structure
as functions of $s$, as a consequence unwanted poles are introduced.
Such poles spoil analyticity, hence they must be subtracted by hand or, equivalently, 
using the DR procedure defined in Appendix~A.\\
More in detail, for both BW formulae we have only one real pole,
that we call $s_s$ and $s_1$ respectively (both less than $\tilde{s}_0$). The corresponding residues,
that we call $R_{s,1}$, are 
\be
\begin{array}{l}
\displaystyle R_s=\frac{s_s}{M^2-2s_s+\frac{3}{2}\gamma_s\sqrt{\tilde s_0-s_s}}\,,\\
\\
\displaystyle R_1=\frac{1}{-1+\frac{3}{2}\gamma_1\sqrt{\tilde s_0-s_1}}\,.\\
\end{array}
\label{eq:residues}
\en
Following eq.~(\ref{eq:regu0}), the regularized BW formulae read
\be
\widetilde{BW}_{s,1}(s)=BW_{s,1}(s)-\frac{R_{s,1}}{s-s_{s,1}} \,.
\no
\en
In particular, below the threshold $\tilde s_0$, where BW's are real, we have
\be
\begin{array}{l}
\displaystyle 
\widetilde{BW}_{s}(s<\tilde s_0)=\frac{s}{s(M^2-s)-\gamma_s(\tilde s_0-s)^{3/2}}
-\frac{R_{s}}{s-s_{s}} \,,\\
\\
\displaystyle
\widetilde{BW}_{1}(s<\tilde s_0)=\frac{1}{M^2-s-\gamma_1(\tilde s_0-s)^{3/2}}
-\frac{R_{1}}{s-s_{1}} \,.\\
\end{array}
\label{eq:bw-below}
\en 
Above $\tilde s_0$ BW's become complex, real and imaginary parts are obtained as
limit of $\widetilde BW_{s,1}(s)$ over the upper edge of the cut $(\tilde s_0,\infty)$.
Since the poles $s_{s,1}$ are real only the real parts have to be corrected as
\be
\begin{array}{l}
\displaystyle 
\re\left[\widetilde{BW}_{s}(s>\tilde s_0)\right]=\frac{s^2(M^2-s)}{s^2(M^2-s)^2+
\gamma_s^2(s-\tilde s_0)^3}
-\frac{R_{s}}{s-s_{s}} \,,\\
\\
\displaystyle 
\re\left[\widetilde{BW}_{1}(s> \tilde s_0)\right]=\frac{M^2-s}{(M^2-s)^2+\gamma_1^2(s-\tilde s_0)^3}
-\frac{R_{1}}{s-s_{1}} \,,\\
\end{array}
\label{eq:bwregu-re}
\en
while the imaginary parts remain unchanged
\be
\begin{array}{l}
\displaystyle 
\im\left[\widetilde{BW}_{s}(s>\tilde s_0)\right]=\frac{s\gamma_s(s-\tilde s_0)^{3/2}}{s^2(M^2-s)^2+
\gamma_s^2(s-\tilde s_0)^3}\,,\\
\\
\displaystyle
\im\left[\widetilde{BW}_{1}(s>\tilde s_0)\right]=\frac{\gamma_1(s-\tilde s_0)^{3/2}}
{(M^2-s)^2+\gamma_1^2(s-\tilde s_0)^3}\,.\\
\end{array}
\label{eq:bwregu-im}
\en
\begin{table}[h!]
\bc
\renewcommand{\arraystretch}{1.1}
\begin{tabular}{l||c|c|c|c|c}
Resonance & $M$ (GeV) &  $\Gamma_0$ (GeV) & $\tilde s_0$ & $s_s$ (GeV$^2$) & $s_1$ (GeV$^2$) \\
\hline\hline
$\rho(770)$ & 0.7755 & 0.1491 & $4M_\pi^2$ & 0.005953 & -11.63 \\
\hline
$\rho(1450)$ & 1.465 & 0.400 & $4M_\pi^2$ & 0.003969 &  -29.43 \\
\hline
$\omega(1420)$ & 1.425 & 0.215 & $(M_\pi+\mr)^2$ & 0.06239 & -19.46\\
\end{tabular}
\caption{\label{tab:param}Parameters for the BW formulae of resonances: $\rho(770)$, $\rho(1450)$ 
and $\omega(1420)$.}
\ec
\end{table}\\
The parameters of the subtracted poles for the three vector mesons are reported 
in Table~\ref{tab:param}.
A third case is discussed in Appendix~B. It is not fitted to the data because 
its resonance structure is intermediate between the two above cases.
%
%
% PERHAPS THE FOLLWING SUBSECTION SHOULD GO INTO AN APPENDIX
% AS IT IS NOT USED IN THE FITS. DONE!
%
%%
\bfi[h]
\bm{75mm}
\epsfig{file=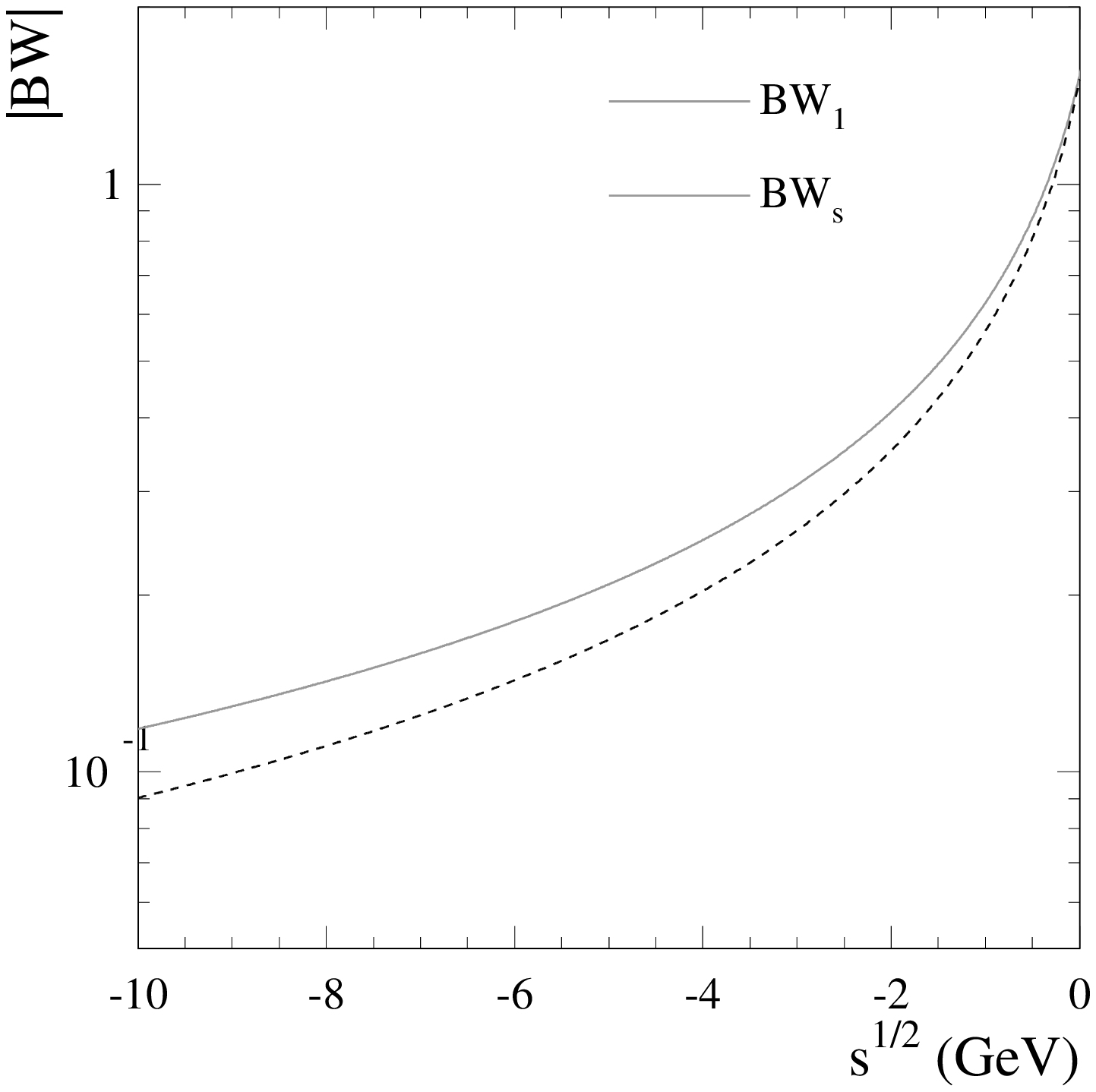,width=75mm}
\em\hfill
\bm{75mm}
\epsfig{file=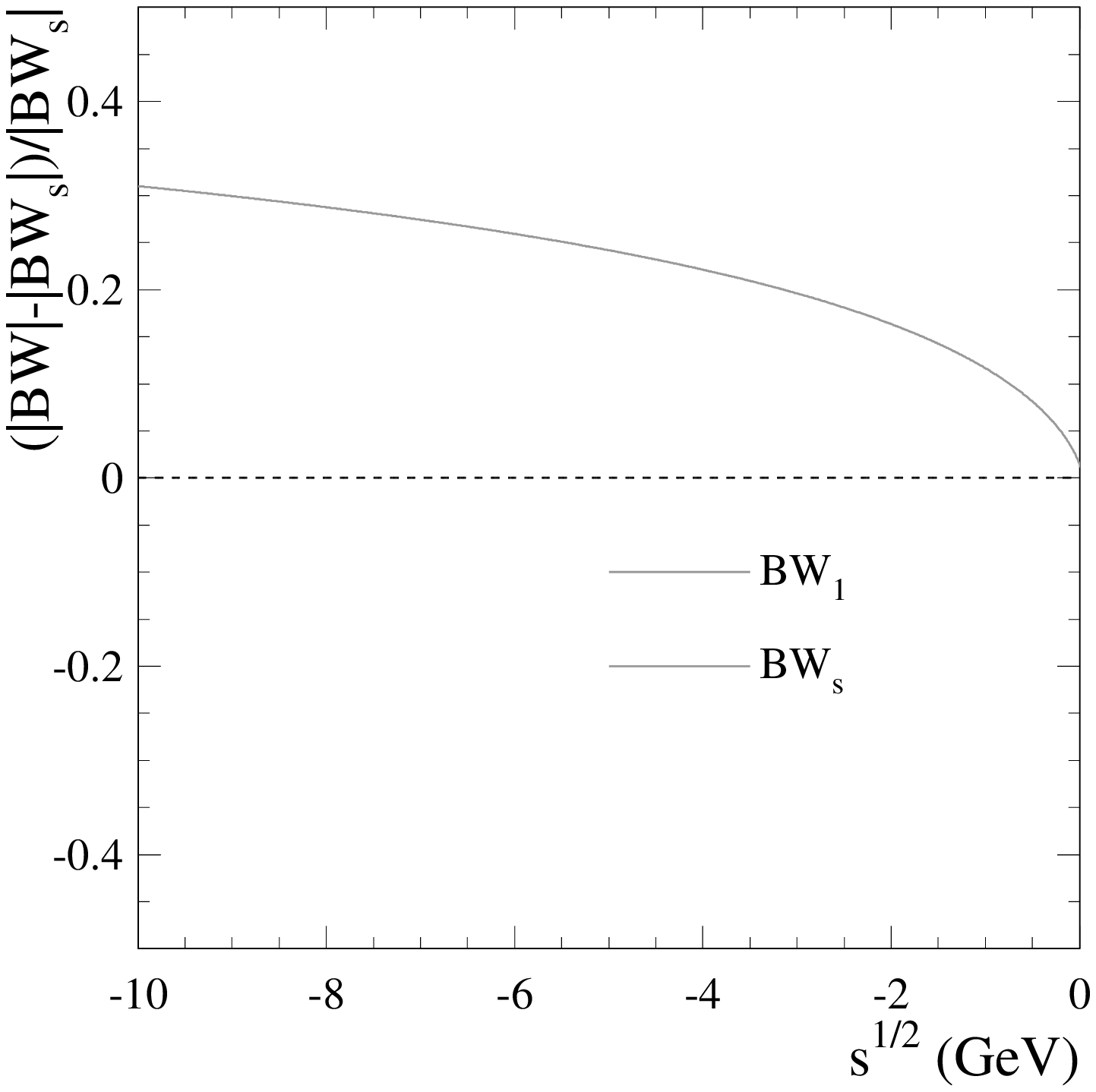,width=75mm}
\em
\caption{Left: comparison between the two descriptions of the $\rho$ 
peak in the space-like region. Right: relative differences w.r.t. $BW_s$.
\label{fig:confsl}}
\efi
\bfi[h]
\bm{75mm}
\epsfig{file=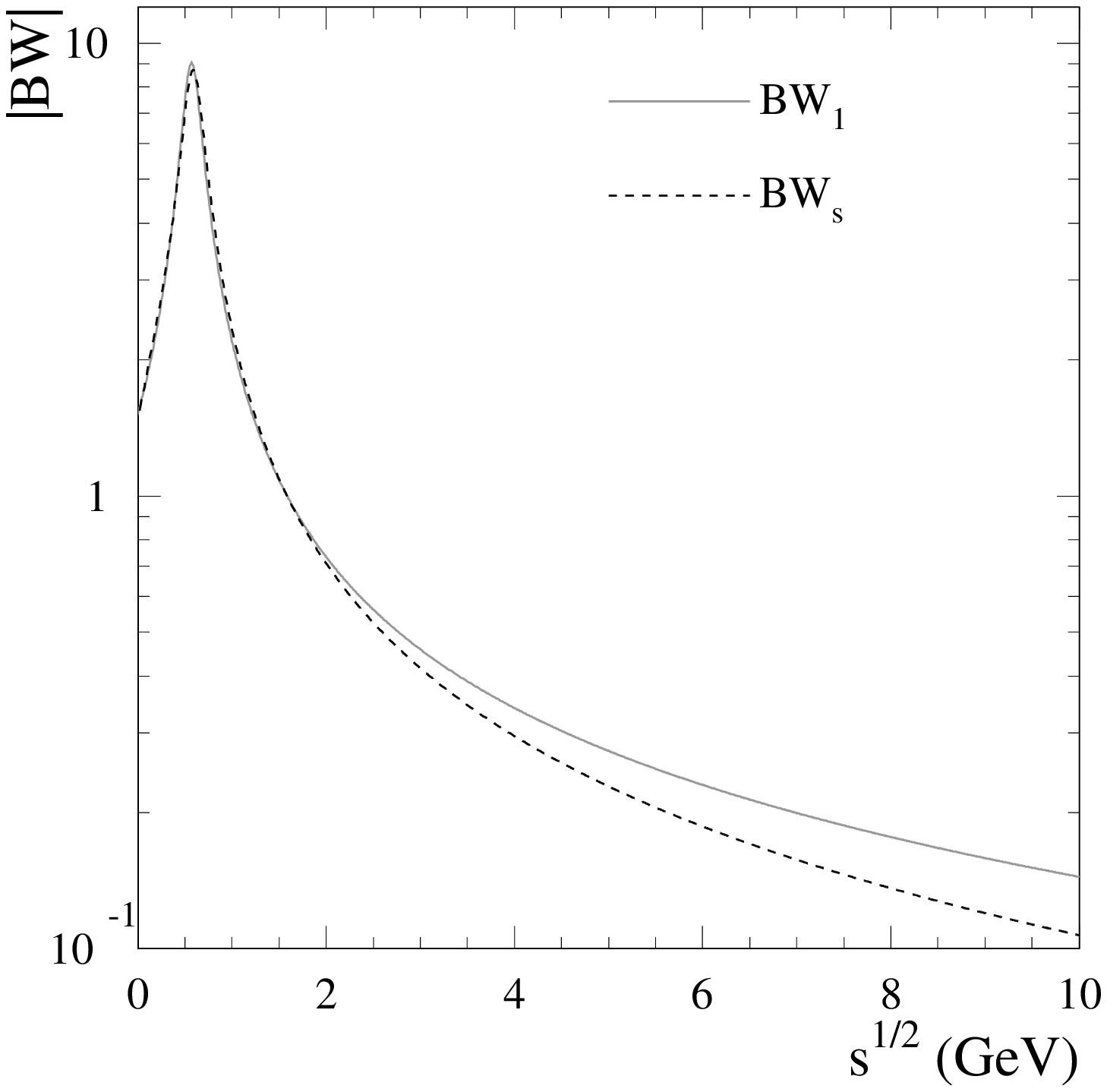,width=75mm}
\em\hfill
\bm{75mm}
\epsfig{file=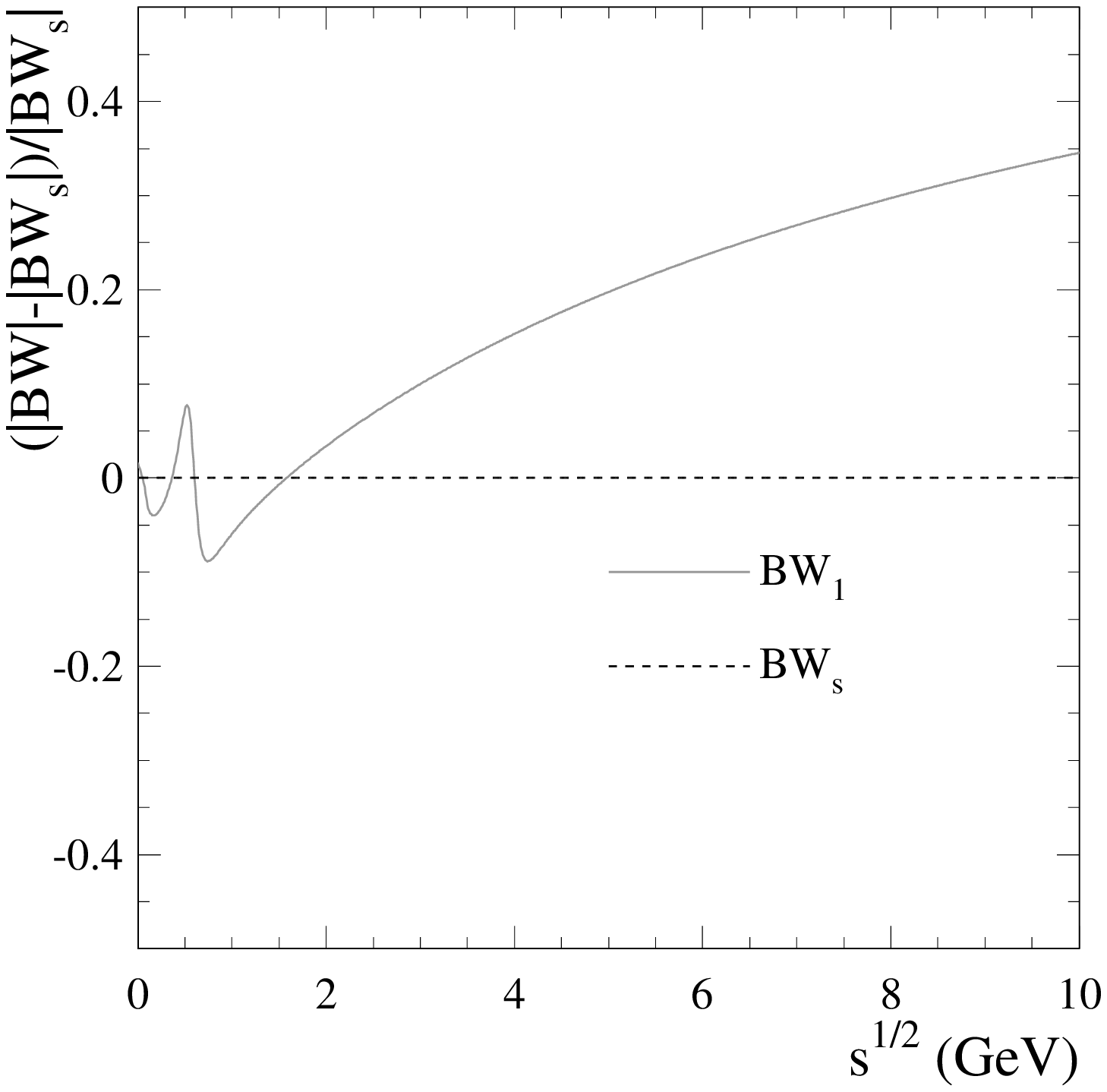,width=75mm}
\em
\caption{Left: comparison between the two descriptions of the $\rho$ 
peak in the time-like region. Right: relative differences w.r.t. $BW_s$.
\label{fig:conf}}
\efi\\
%
%  included in January 2012
%
Figures~\ref{fig:confsl} and~\ref{fig:conf} show comparisons between 
the two descriptions in case of $\rho$ in the space-like and
time-like regions respectively. On the left of each figure
is the modulus of each, on the right the relative difference 
with respect to $BW_s$.
\section{The analytic extension}
\label{sec:extension}
The original model, described in Sec.~\ref{sec:model} and constructed
in the space-like region, can be analytically continued in the time-like 
region using the regularized BW formulae obtained in Sec.~\ref{sec:analytic}.
We consider then a new set of expressions for $\fiv_{1,2}(q^2)$ and 
$\fis_{1,2}(q^2)$, homologous to those of eq.~(\ref{eq:gk-ff}) where now we 
use regularized BW formulae instead of the MMD~\cite{meissner} $\rho$ width form or the 
zero-width approximation given in eq.~(\ref{eq:bw0}), and also two additional 
vector meson contributions, $\rho(1450)$ and $\omega(1420)$ here simply $\rho'$ 
and $\omega'$, as in the last version of the Lomon model~\cite{lomon-list}. 
Such BW's have the expected analytic structure and reproduce in both 
space-like and time-like regions the finite-width effect of broad resonances.
The narrow widths of the $\omega$ and $\phi$ have negligible effects, so we use 
these modified propagators only for broader 
vector mesons, namely: the isovectors $\rho$ and $\rho'$, and the isoscalar 
$\omega'$. These are the new expressions for the isospin components of 
nucleon EMFF's
\be
\begin{array}{rcl}
\fiv_{1,\rm case}(q^2)&=&\ds \big[\widetilde{BW}_{\rm case}^\rho(q^2)
+\widetilde{BW}_{\rm case}^{\rho'}(q^2)\big]F_1^\rho(q^2)\vspace{2mm}\\
&&\ds+\big[1-\widetilde{BW}_{\rm case}^{\rho}(0)-\widetilde{BW}_{\rm case}^{\rho'}(0)\big]F_1^D(q^2)\\
&&\\
\fiv_{2,\rm case}(q^2)&=&\ds \big[\kappa_\rho\,\widetilde{BW}_{\rm case}^\rho(q^2)
+\kappa_{\rho'}\,\widetilde{BW}_{\rm case}^{\rho'}(q^2)\big]F_2^\rho(q^2)\vspace{2mm}\\
&&\ds+\big[\kappa_{\rm v}-
\kappa_{\rho}\,\widetilde{BW}_{\rm case}^{\rho}(0)-
\kappa_{\rho'}\,\widetilde{BW}_{\rm case}^{\rho'}(0)\big]F_2^D(q^2)\\
&&\\
\fis_{1,\rm case}(q^2)&=&\ds\big[ BW_0^\omega(q^2)+\widetilde{BW}_{\rm case}^{\omega'}(q^2)\big]
F_1^\omega(q^2)+BW_0^\phi(q^2)\,F_1^\phi(q^2)+\vspace{2mm}\\
&&\big[1-BW_0^\omega(0)-\widetilde{BW}_{\rm case}^{\omega'}(0)\big]F_1^D(q^2)\\
&&\\
\fis_{2,\rm case}(q^2)&=&\ds\big[ \kappa_\omega\,BW_0^\omega(q^2)
+\kappa_{\omega'}\,\widetilde{BW}_{\rm case}^{\omega'}(q^2)\big]F_2^\omega(q^2)+
\kappa_\phi\,BW_0^\phi(q^2)\,F_2^\phi(q^2)+\vspace{2mm}\\
&&\big[\kappa_{\rm s}-\kappa_\omega\,BW_0^\omega(0)-
\kappa_{\omega'}\,\widetilde{BW}_{\rm case}^{\omega'}(0)-
\kappa_\phi\,BW_0^\phi(0)\big]F_2^D(q^2)\,,\\
\no\label{eq:gk-lomon}
\end{array}
\en
where case=$s$ and case=1 correspond to the parametrizations of 
the energy dependent width described in Sec.~\ref{subsec:bws}.
%while case\,=\,$\sqrt{s}$ is the third possibility treated in 
%Sec.~\ref{subsec:third}. 
Following eqs.~(\ref{eq:bw-below})-(\ref{eq:bwregu-im}) 
%and eq.~(\ref{eq:3case}) 
for the definition of $\widetilde{BW}(q^2)$, and including 
the coupling constants, we have
\be
\widetilde{BW}^{\beta}_{\rm case}(q^2)=
\left\{\begin{array}{lcl}
\displaystyle 
\frac{g_\beta M^2_\beta}{f_\beta}
\left[\frac{q^2}{q^2(M^2_\beta-q^2)-i\gamma_s^\beta(q^2-\tilde s_0^\beta)^{3/2}}
-\frac{R_{s}^\beta}{q^2-s_{s}^\beta}\right] &\hspace{5mm}& {\rm case}=s\\
&&\\
\displaystyle
\frac{g_\beta M^2_\beta}{f_\beta}
\left[\frac{1}{M^2_\beta-q^2-i\gamma_1^\beta(q^2-\tilde s_0^\beta)^{3/2}}
-\frac{R_{1}^\beta}{q^2-s_{1}^\beta}\right]  && {\rm case}=1\\
%&&\\
%\displaystyle
%\frac{g_\beta M^2_\beta}{f_\beta}
%\widetilde D(q^2)  && {\rm case}=\sqrt{s}\,,\\
\end{array}\right.
\no\label{eq:bwregu-tot}
\en
with: $\beta=\rho$, $\rho'$, $\omega'$ (parameters in Table~\ref{tab:param})
and where: the $\gamma_{1,s}^\beta$ are given in eq.~(\ref{eq:widths})
and the residues $R_{1,s}^\beta$ in eq.~(\ref{eq:residues}).
The introduction of the regularized BW's does not spoil the high-energy
behavior of the resulting nucleon EMFF's. In fact, as $|q^2|\to\infty$, the function 
$\widetilde{BW}^\beta_{\rm case}(q^2)$ vanishes like $1/q^2$, i.e. following the same power law 
as the previous $BW^\beta_{0}(q^2)$, in the case=1 and case=$s$, 
%while in case\,=\,$\sqrt{s}$ it goes like $\ln|q^2|/q^2$, 
see eq.~(\ref{eq:bw0}),
% and~(\ref{eq:asy3}), 
indeed we have
\be
\widetilde{BW}^\beta_{\rm case}(q^2)\mathop{\sim}_{|q^2|\to\infty}
\left\{\begin{array}{lcl}
\ds\frac{g_\beta M^2_\beta}{f_\beta}\,
\frac{1-R_s^\beta}{q^2} &\hspace{5mm}& {\rm case}=s\\
&&\\
\ds-\frac{g_\beta M^2_\beta}{f_\beta}\,\frac{R_1^\beta}{q^2} && {\rm case}=1\\
%&&\\
%\ds\frac{g_\beta M^2_\beta}{f_\beta}\,\frac{\gamma_0}{\pi(1+\gamma_0^2)}\,
%\frac{\ln|q^2|}{q^2}&& {\rm case}=\sqrt{s}\,.\\
\end{array}\right.\,.
\no\en
It is interesting to notice that in both cases is just the subtracted 
pole which ensures the expected behavior and, in particular, the asymptotic
limit of: $q^2\cdot\widetilde{BW}^\beta_{\rm case}(q^2)$ is proportional to
$R^\beta_1$ and $(1-R^\beta_s)$ respectively.
For the reason discussed at the end of Sec.~\ref{sec:model}, for the present 
model $q$ is replaced by $\tilde q$ in the hadronic FF's $F_i^{\phi}$ of eq.~(\ref{eq:fi}). 
\section{Results}
\label{sec:results}
Nine sets of data have been considered, six of them lie in the
space-like region~\cite{SL-data} and three in the 
time-like region~\cite{fenice,dm1,dm2,bes,cleo,lear,e760,e835,babar}. 
The data determine the Sachs EMFF's and their ratios. The fit procedure
consists in defining a global $\chi^2$ as a sum of nine
contributions, one for each set. More in detail, we minimize
the quantity
\be
\chi^2=\sum_{i=1}^9\tau_i\cdot\chi^2_{i}\,,
\no\label{eq:chi2tot}
\en
where the coefficients $\tau_i$ weight the $i^{\rm th}$ contribution,
we use $\tau_i=1$ or $\tau_i=0$ to include or exclude the 
$i^{\rm th}$ data set. The single contribution, $\chi^2_{i}$, is defined in
the usual form as
\be
\chi^2_{i}=\sum_{k=1}^{N_i}\left(\frac{Q_i(q^2_k)-v^i_k}{\delta v^i_k}\right)^2\,,
\no\en
where $Q_i(q^2)$ indicates the physical observable, function of $q^2$, that 
has been measured and the set $\{q^2_k,v^i_k,\delta v^i_k;N_i\}$ represents the 
corresponding data; $v^i_k$ is the $k^{\rm th}$ value ($k=1,\ldots,N_i$)
of the quantity $Q_i$ ($i=1,\ldots,9$) measured at $q^2=q^2_k$, with  
error $\delta v^i_k$.
%
%Observables
%
\begin{table}[h!]
\bc
\renewcommand{\arraystretch}{1.7}
\begin{tabular}{|c|c|r|r|r|r|r|}
\cline{2-7}
\multicolumn{1}{c|}{} 
& \multirow{2}{*}{$Q_i$} & \multirow{2}{*}{$N_i$} & \multicolumn{4}{|c|}{ minimum $\chi^2_i$} \\
\cline{4-7}
\multicolumn{1}{c|}{} & & 
& \bm{22mm}\vspace{1mm}\centering case\,=\,$s$\\ With \bbr\vspace{1mm}\em
& \bm{22mm}\vspace{1mm}\centering case\,=\,1\\ With \bbr\vspace{1mm}\em
& \bm{22mm}\vspace{1mm}\centering case\,=\,$s$\\ No \bbr\vspace{1mm}\em
& \bm{22mm}\vspace{1mm}\centering case\,=\,1\\ No \bbr\vspace{1mm}\em
\\
\hline\hline
\multirow{6}{*}{\rotatebox{90}{space-like}} 
& $G_M^p$ & 68 & 48.7 & 50.1 & 54.6 & 60.8\\
\cline{2-7}
& $G_E^p$ & 36 & 30.4 & 27.6 & 26.2 & 35.0\\
\cline{2-7}
& $G_M^n$ & 65 &  154.6 & 154.2 & 158.2 & 167.0\\
\cline{2-7}
& $G_E^n$ & 14 & 22.7 & 23.2 & 24.1 & 26.0\\
\cline{2-7}
& $\mu_p G_E^p/G_M^p$ & 25 & 13.9 & 12.9 & 10.6 &14.4\\
\cline{2-7}
& $\mu_n G_E^n/G_M^n$ & 13 & 11.3 & 10.7 & 8.2 & 8.9\\
\hline\hline
\multirow{2}{*}{\rotatebox{90}{\!\!time-like}}
& $|G_{\rm eff}^p|$ & 81 (43) & 162.5 & 166.7 & 62.2 & 35.0\\
\cline{2-7}
& $|G_{\rm eff}^n|$ & 5 & 8.4 & 6.3 & 3.2 & 0.3\\
%\cline{2-7}
%& $|G_E^p/G_M^p|$ & 6 & 9.98 & 10.28 &&\\
\hline\hline
\multicolumn{1}{c|}{} & Total& 313(275) & 452.5 & 451.7 & 347.3 & 347.4\\
\cline{2-7}
\end{tabular}
\caption{\label{tab:chi2-cont}%
Measured quantities, numbers of data points and $\chi^2$ contributions.
The values in parentheses indicate the number of data points in the case 
``No \bbr''.}
\ec
\end{table}\\
Table~\ref{tab:chi2-cont} reports the complete list of observables, the 
number of data points and the corresponding minimum $\chi^2$'s, in the 
two considered cases as described in Sec.~\ref{sec:extension} 
for the sets of data with and without the \bbr\ data which have 
a final state photon.  For case=$s$,  with and without the \bbr\ data, the 
optimization over the full set of 13 free parameters (Table~\ref{tab:best-parms}) 
determines $\Lambda_1$, $\Lambda_2$, $\Lambda_D$ and $\Lambda_{\rm QCD}$ 
such that the hadronic FF's have no real poles.  For the case=$1$ with 
\bbr\ data the full minimization implies a zero for $(\Lambda_1^2- \tilde q^2)$ producing 
poles in the hadronic FF.  Re-minimizing with the constraint $\Lambda_1=0.5\,\gev$, 
just above the $0.4744\,\gev$ obtained without the constraint, removes the poles.  
For case=$1$ without \bbr\ data it is required that the already fixed 
$\Lambda_{\rm QCD}=0.15\,\gev$ be changed to $\Lambda_{\rm QCD}=0.10\,\gev$  
to avoid a zero of $(\Lambda_1^2- \tilde q^2)$.  In both cases the change in 
$\chi^2$ is negligible.
\\
Data and fits, black and gray curves correspond to case=1 and case=$s$ 
respectively, are shown in Figs.~\ref{fig:gep}-\ref{fig:rptl}. In the space-like
region the electric Sachs EMFF's are normalized to the dipole form
\be
G_D(q^2)=\left(1-\frac{q^2}{0.71\,{\rm GeV}^2}\right)^{-2}\,,\no
\en
while magnetic EMFF's are also normalized to the magnetic moment.
This normalization decreases the range of variation, but the curves 
clearly demonstrate deviations from the dipole form.
The observable $R_N$ is defined as the ratio $R_N=G_E^N/G_M^N$ 
for the nucleon $N$.   As $N$ stands for both neutron and proton 
there are six space-like observables.
A departure from scaling is shown in the deviation of $R_p$ and $R_n$ from unity.
\\
The time-like effective FF, $|G_{\rm eff}^N|$, is defined as 
\be
|G_{\rm eff}^N(q^2)|=\left[\frac{\sigma(\ee\to\NN)}{
\frac{4\pi\alpha^2}{3q^2}\sqrt{1-\frac{4M_N^2}{q^2}}
\left(1+\frac{2M_N^2}{q^2}\right)}\right]^{1/2}\,,
\label{eq:geff-data}
%\no
\en
where $\sigma(\ee\to\NN)$ is the measured total cross section and
the kinematic factor at denominator is the Born cross section 
for a pointlike nucleon.
In terms of electric and magnetic EMFF's, $G_E^N$ and $G_M^N$, 
i.e. considering the matrix element given in eq.~(\ref{eq:Gamma})
and the definitions of eq.~(\ref{eq:sachs}),
we have
\be
|G_{\rm eff}^N(q^2)|=\left(\big|G_M^N(q^2)\big|^2+
\frac{2M_N^2}{q^2}\big|G_E^N(q^2)\big|^2\right)^{1/2}
\left(1+\frac{2M_N^2}{q^2}\right)^{-1/2}\,,
\label{eq:geff-fit}
\no
\en
and this is the relation that we use to fit the data on $|G_{\rm eff}^N|$ 
for both proton-antiproton and neutron-antineutron production.
\\
The proton-antiproton production experiments were of two types, 1) 
the exclusive pair production~\cite{fenice,dm1,dm2,bes,cleo,lear,e760,e835} and 2) production of 
the pair with a photon~\cite{babar}.  In the latter case the pair 
production energy is obtained by assuming that the photon was 
produced by the electron or positron and that no other photons were 
emitted but undetected.  Fits of the model were made both with and 
without the latter data~\cite{babar}. In Figs.~\ref{fig:gep}-\ref{fig:rptl}
the fit curves corresponding to the two possibilities: with
and without \bbr\ data, are shown as 
solid and dashed lines, respectively.
\\
The free parameters of this model are:
\bi
\item the three cut-offs: $\Lambda_1$, $\Lambda_2$ and $\Lambda_D$ which parameterize 
      the effect of hadronic FF's and control the 
      transition from non-perturbative to perturbative QCD regime in the
      $\gamma NN$ vertex;
\item five pairs of vector meson anomalous magnetic moments and photon couplings
      $(\kappa_\alpha,g_\alpha/f_\alpha)$, with $\alpha=\rho$, $\rho'$,
      $\omega$, $\omega'$, $\phi$.
\ei
The best values for these 13 free parameters together with the constants of this model
are reported in Table~\ref{tab:best-parms}.\\
The fixed parameters concern well known measurable features of
                the intermediate vector mesons and dynamical quantities. 
		Particular attention has to be paid to $\Lambda_{\rm QCD}$.
                In fact we use the values $\Lambda_{\rm QCD}=0.15$ GeV in all cases
                but for the case=1 without \bbr\ data, where instead: 
                $\Lambda_{\rm QCD}=0.10$~GeV.
                The use of such a reduced value is motivated by the requirement of having no
                real poles in meson-nucleon and quark-nucleon FF's (Sec.~\ref{sec:model}).
As $\Lambda_{\rm QCD}=0.15\,\gev$ is closer to the values preferred by high energy experiments,
it suggests that case=$s$ is the more physical model. Another reason to prefer it 
on physical grounds is that the width formula of the vector meson decay in case=s 
is determined by relativistic perturbation theory.  Case=1 was chosen because it is 
a simpler relativistic modification of the non-relativistic Breit-Wigner form.  
This in our view is a less physical reason.
%
% Best paramters
%
\begin{table}[h!]
\bc
\renewcommand{\arraystretch}{1.3}
\begin{tabular}{|c|c|c|c|c|}
\hline
Parameter
& \bm{22mm}\vspace{1mm}\centering case\,=\,$s$\\ With \bbr\vspace{1mm}\em
& \bm{22mm}\vspace{1mm}\centering case\,=\,1\\ With \bbr\vspace{1mm}\em
& \bm{22mm}\vspace{1mm}\centering case\,=\,$s$\\ No \bbr\vspace{1mm}\em
& \bm{22mm}\vspace{1mm}\centering case\,=\,1\\ No \bbr\vspace{1mm}\em
\\
\hline\hline

$g_\rho/f_\rho$ &2.766  &2.410     &0.9029   &0.4181 \\ \hline
$\kappa_\rho$ & -1.194 & -1.084     &0.8267 & 0.6885 \\ \hline
$M_\rho$ (GeV)&\multicolumn{4}{|c|}{0.7755 (fixed)} \\ \hline
$\Gamma_\rho$ (GeV)&\multicolumn{4}{|c|}{0.1491 (fixed)} \\ \hline
\hline
$g_\omega/f_\omega$ &-1.057 &-1.043   &-0.2308 &-0.4894 \\ \hline
$\kappa_\omega$ &-3.240& -3.317& -9.859 & -1.398 \\ \hline
$M_\omega$ (GeV)&\multicolumn{4}{|c|}{0.78263 (fixed)} \\ \hline
\hline
$g_\phi/f_\phi$ &  0.1871&  0.1445&   ­0.0131& ­0.1156\\ \hline
$\kappa_\phi$ & -2.004 & -­3.045 &  37.218 & -0.2613\\ \hline
$M_\phi$ (GeV)&\multicolumn{4}{|c|}{1.019 (fixed)} \\ \hline
$\mu_\phi$ (GeV)&\multicolumn{4}{|c|}{20.0 (fixed)} \\ \hline
\hline
$g_{\omega'}/f_{\omega'}$ & 2.015 & 1.974  & 1.265&  1.649 \\ \hline
$\kappa_{\omega'}$ & ­-2.053& ­-2.010&  ­-2.044& ­-0.6712 \\ \hline
$M_{\omega'}$ (GeV) &\multicolumn{4}{|c|}{1.425 (fixed)} \\ \hline
$\Gamma_{\omega'} (GeV)$ &\multicolumn{4}{|c|}{0.215 (fixed)} \\ \hline
\hline
$g_{\rho'}/f_{\rho'}$ &­-3.475& ­-3.274&  ­-0.8730 &­-0.0369 \\ \hline
$\kappa_{\rho'}$ &­-1.657& -­1.724 & ­-2.832& ­-104.35 \\ \hline
$M_{\rho'}$ (GeV)&\multicolumn{4}{|c|}{1.465 (fixed)} \\ \hline
$\Gamma_{\rho'}$ (GeV)&\multicolumn{4}{|c|}{0.400 (fixed)} \\ \hline
\hline
$\Lambda_1$ (GeV) & 0.4801 & 0.5000 & 0.6474 & 0.6446 \\ \hline
$\Lambda_2$ (GeV) & 3.0536 & 3.0562 & 3.0872 & 3.6719 \\ \hline
$\Lambda_D$ (GeV) & 0.7263 & 0.7416 & 0.8573 & 0.8967 \\ \hline
$\Lambda_{\rm QCD}$ (GeV)&\multicolumn{3}{|c|}{0.150} & 0.100 \\ \hline
\hline
\end{tabular}
\caption{\label{tab:best-parms}%
Best values of fit parameters and constants.}
\ec
\end{table}\\
\bfi[h!]
\bc
\bm{78mm}
\epsfig{file=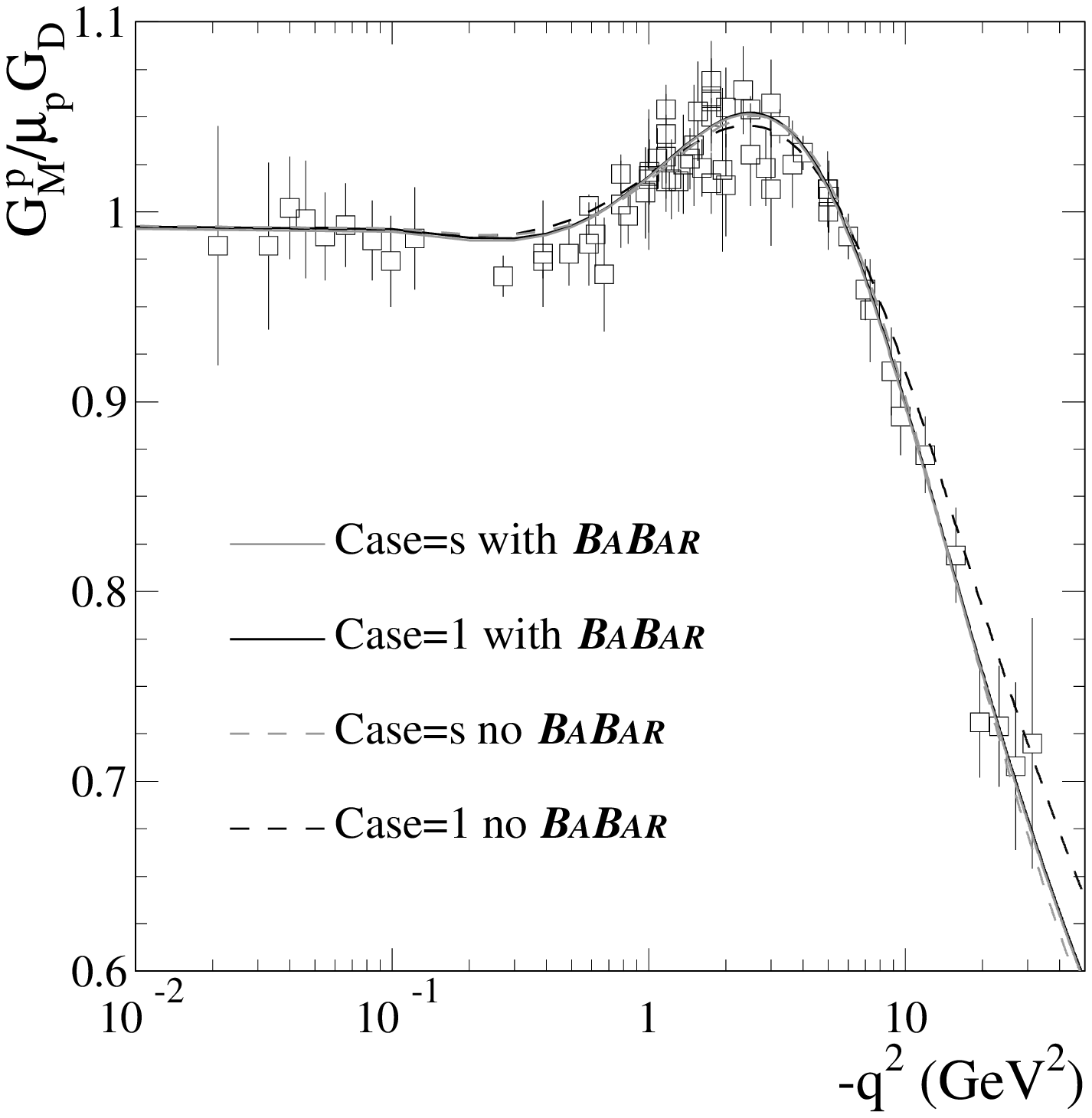,width=78mm}
\caption{\label{fig:gmp} Space-like magnetic proton EMFF  normalized to
the dipole and $\mu_p$, in case=1 and case=$s$, including and not the \bbr\ data.}
\em\hfill
\bm{78mm}
\epsfig{file=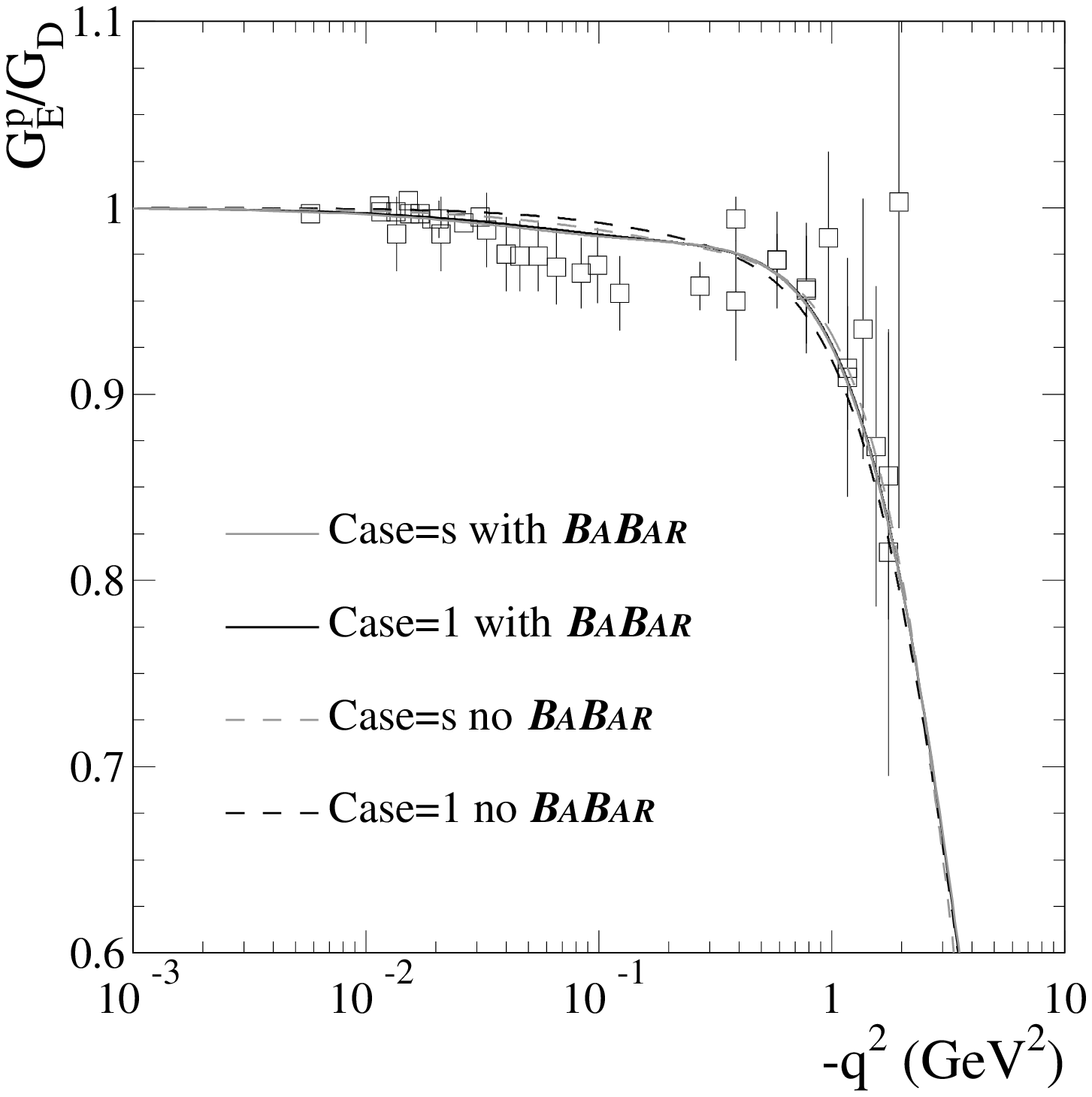,width=78mm}
\caption{\label{fig:gep} Space-like electric proton EMFF normalized to
the dipole, in case=1 and case=$s$, including and not the \bbr\ data.}
\em
\ec
\efi
\bfi[h!]
\bc
\bm{78mm}
\epsfig{file=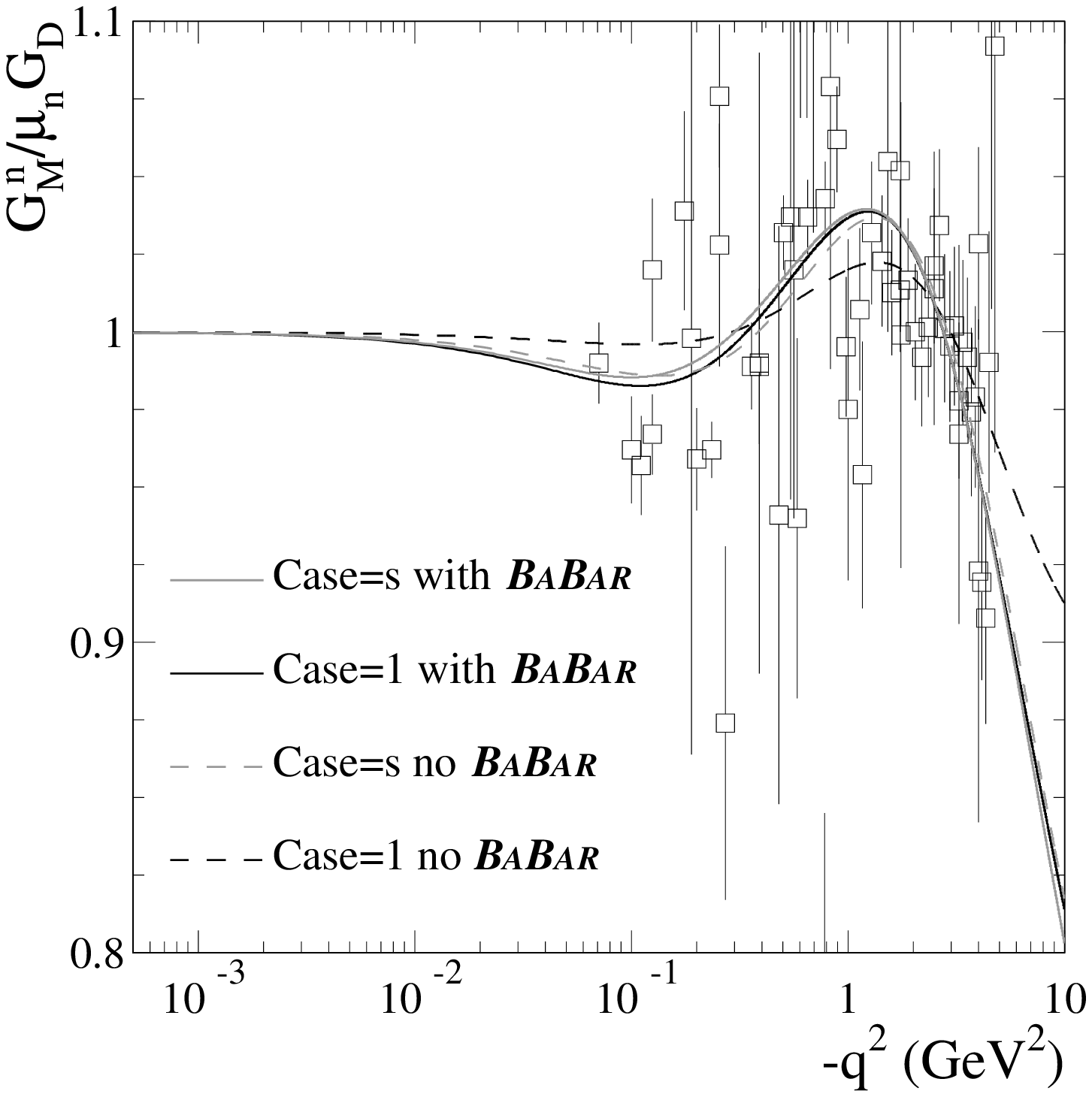,width=78mm}
\caption{\label{fig:gmn} Space-like magnetic neutron EMFF normalized to
the dipole and $\mu_n$, in case=1 and case=$s$, including and not the \bbr\ data.}
\em\hfill
\bm{78mm}
\epsfig{file=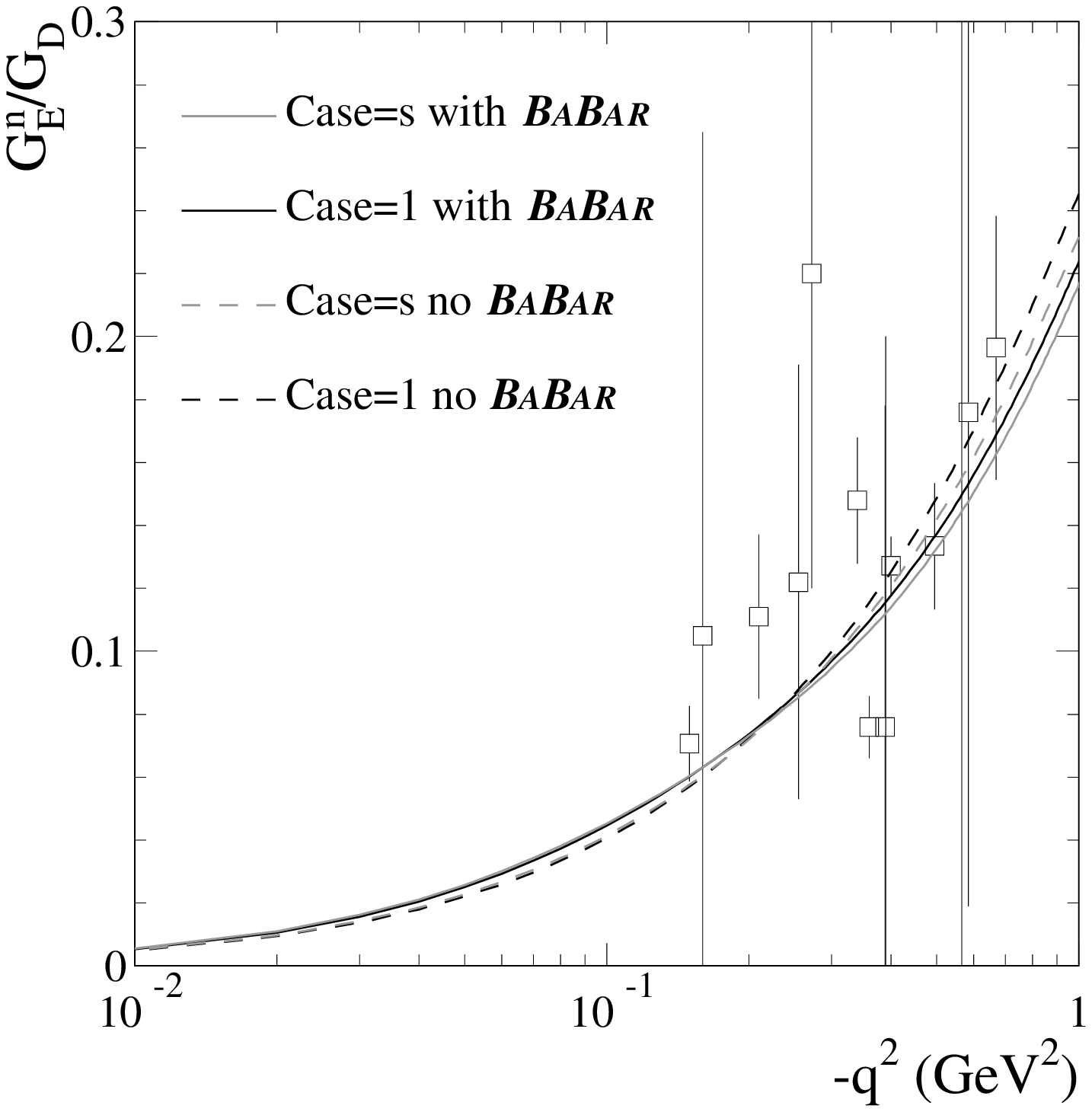,width=78mm}
\caption{\label{fig:gen} Space-like electric neutron EMFF normalized to
the dipole, in case=1 and case=$s$, including and not the \bbr\ data.}
\em
\ec
\efi
\bfi[h!]
\bc
\bm{78mm}
\epsfig{file=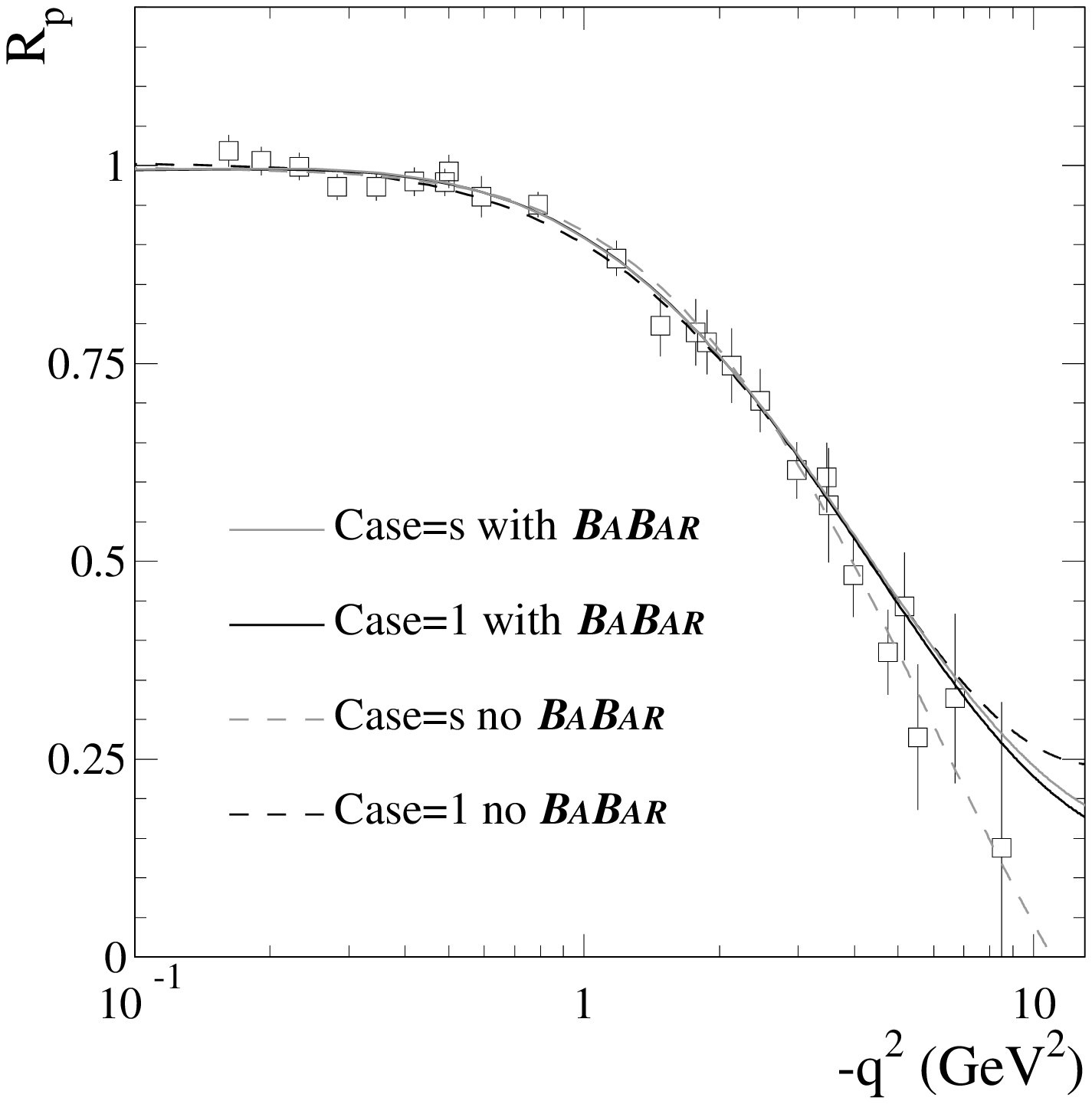,width=78mm}
\caption{\label{fig:rp} Space-like ratio $R_p=\mu_p\,G_E^p/G_M^p$ normalized to
$\mu_p$, in case=1 and case=$s$, including and not the \bbr\ data.}
\em\hfill
\bm{78mm}
\epsfig{file=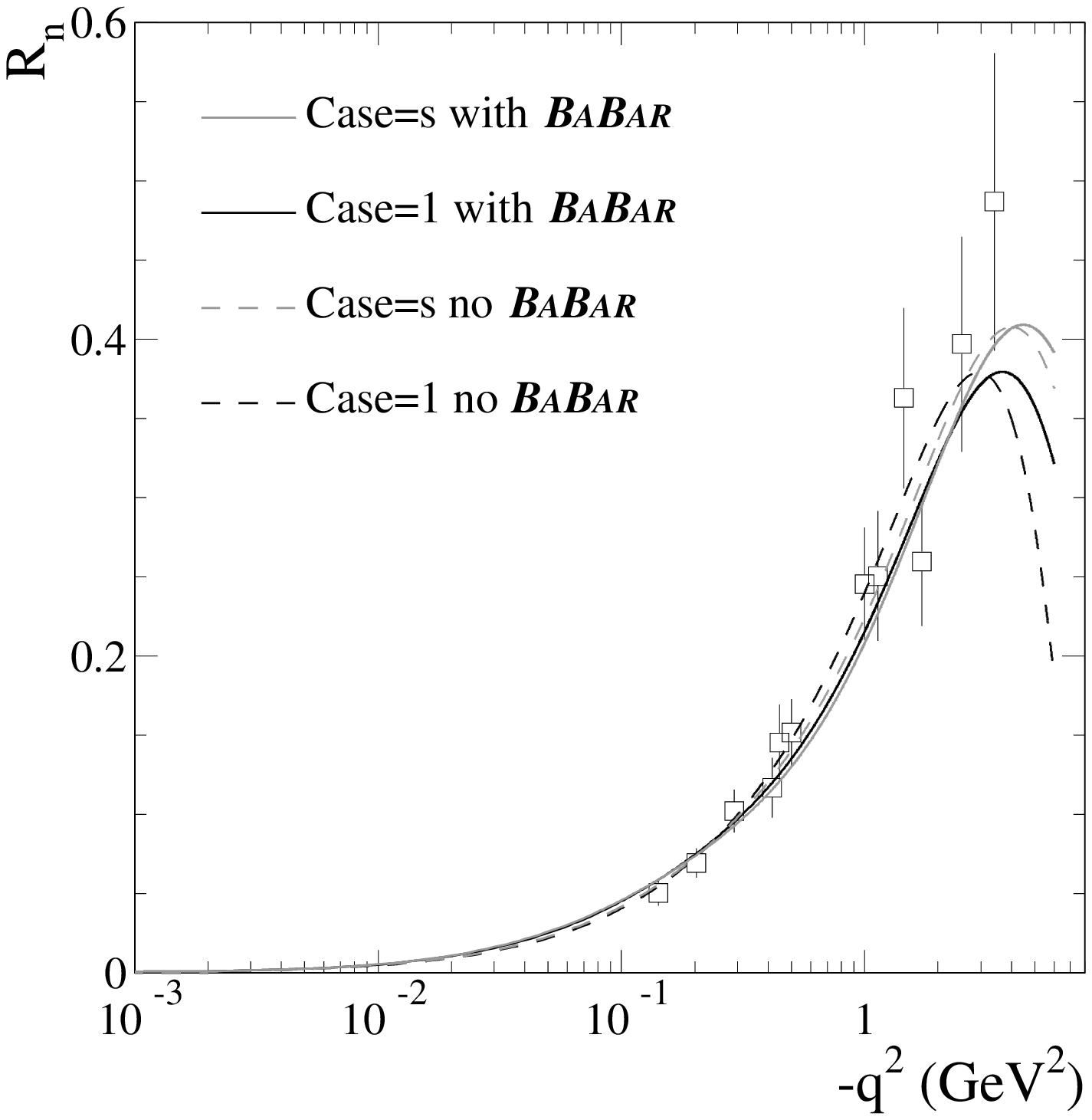,width=78mm}
\caption{\label{fig:rn} Space-like ratio $R_n=\mu_n\,G_E^n/G_M^n$ normalized to
$\mu_n$, in case=1 and case=$s$, including and not the \bbr\ data.}
\em
\ec
\efi
\bfi[h!]
\bc
\bm{78mm}
\epsfig{file=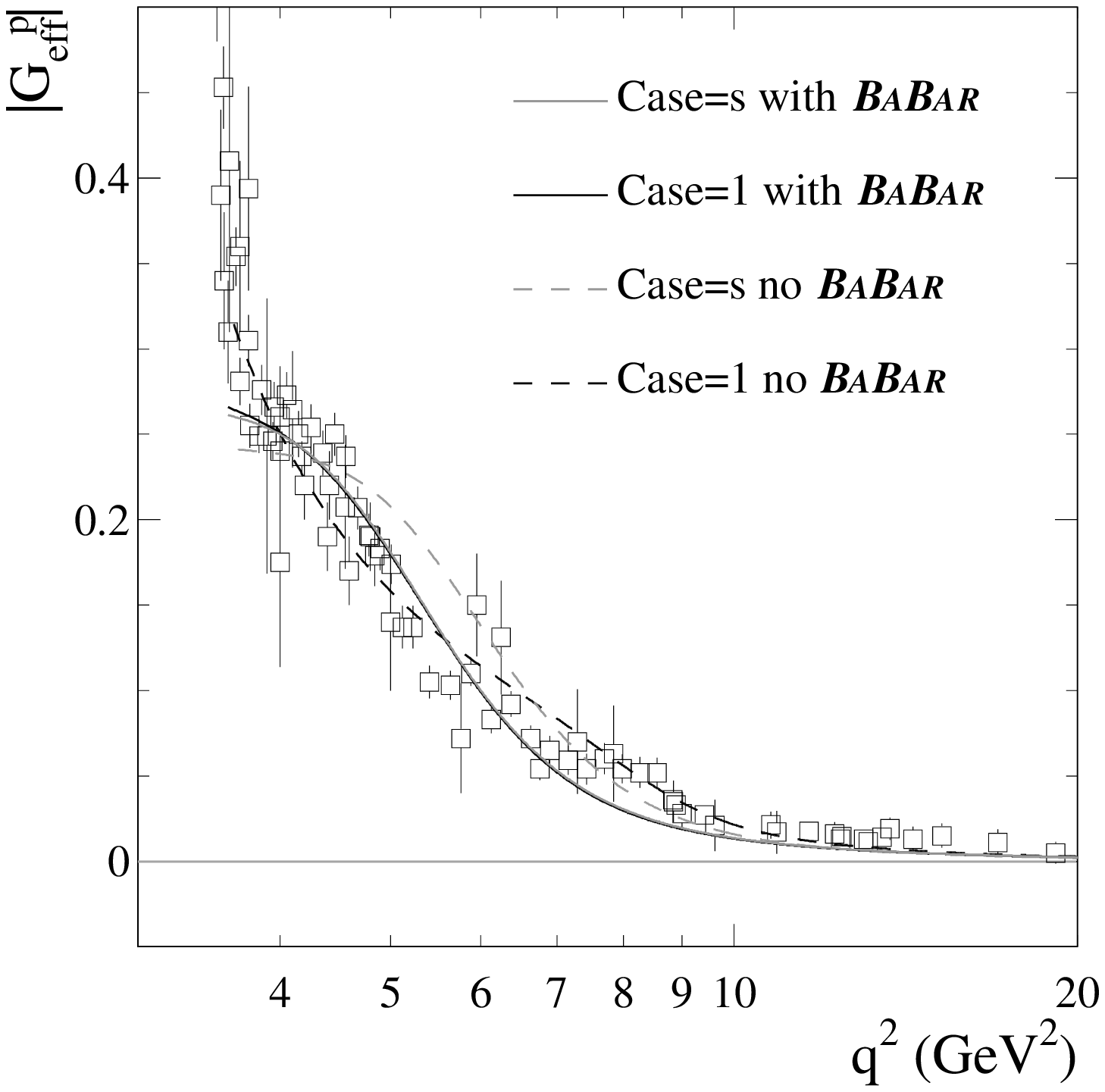,width=78mm}
\caption{\label{fig:geffp} Time-like effective proton FF data
(nine sets~\cite{fenice,dm1,dm2,bes,cleo,lear,e760,e835,babar}) and fit, 
in case=1 and case=$s$, including and not the \bbr\ data.}
\em\hfill
\bm{78mm}
\epsfig{file=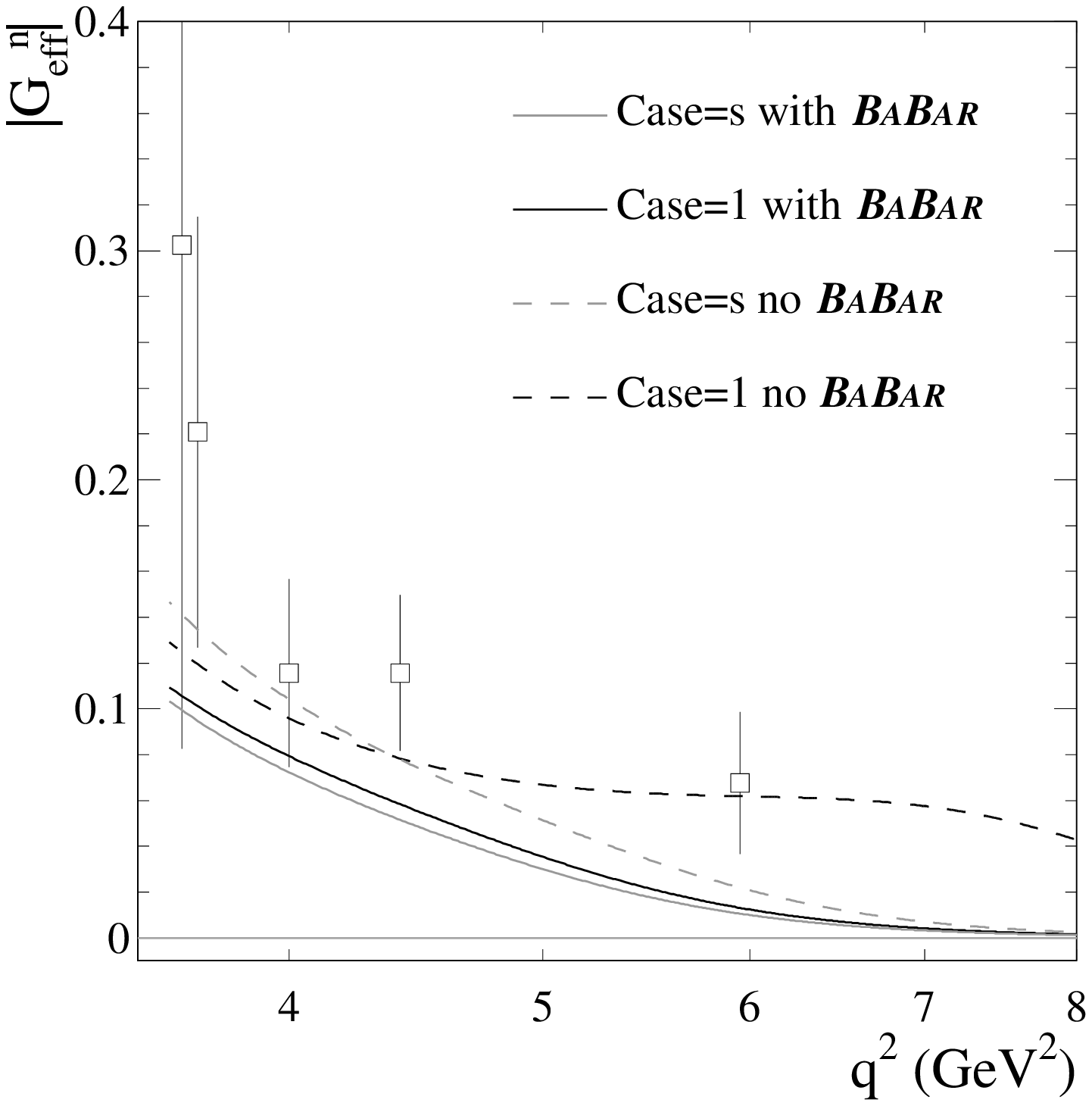,width=78mm}
\caption{\label{fig:geffn} Time-like effective neutron FF data
(only FENICE~\cite{fenice}) and fit, in case=1 and case=$s$, including 
and not the \bbr\ data.}
\em
\ec
\efi
\bfi[h!]
\bc
\bm{78mm}
\epsfig{file=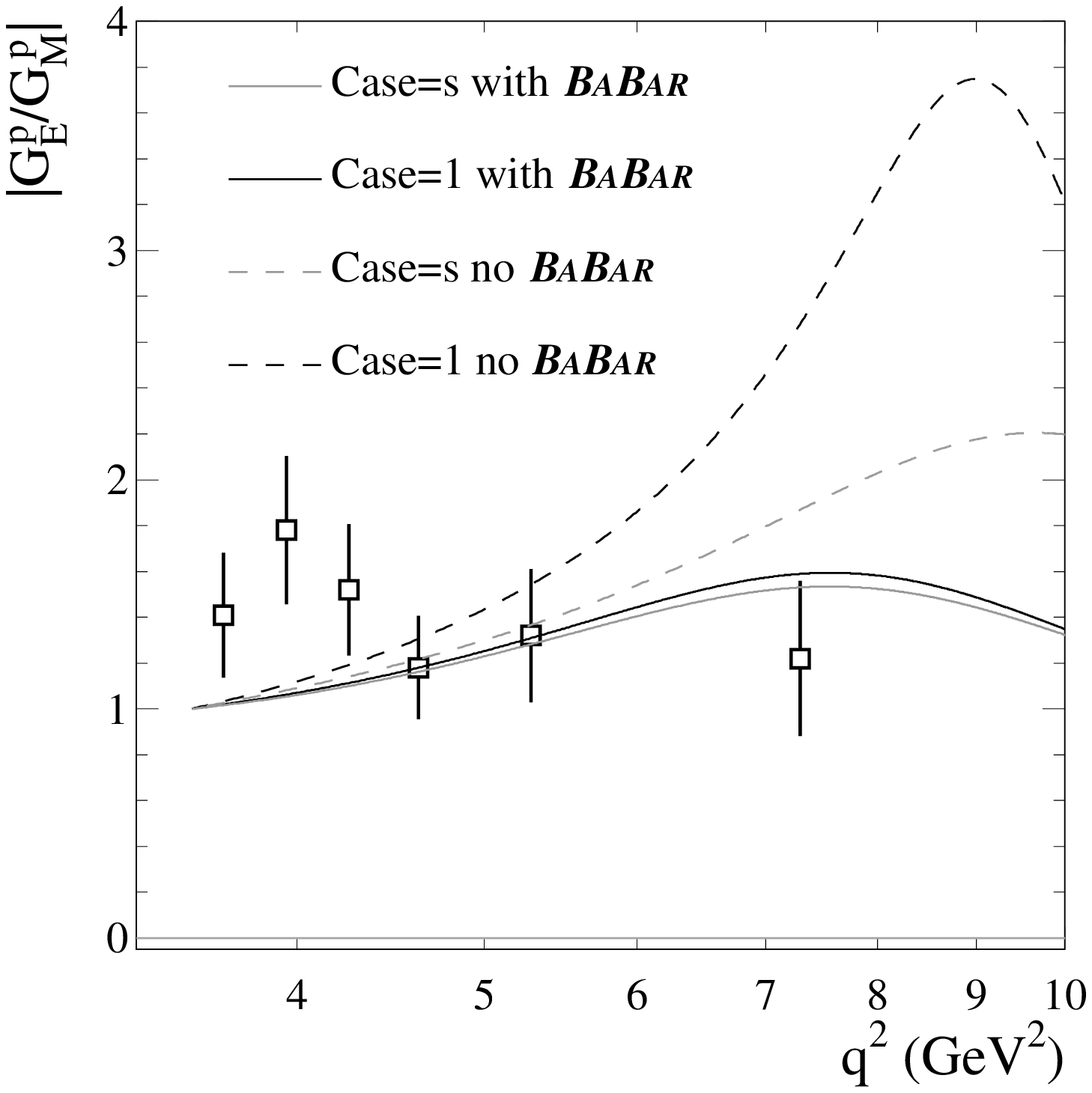,width=78mm}
\caption{\label{fig:rptl} Modulus of the ratio $G_E^p/G_M^p$, data~\cite{babar}
and prediction, in the time-like 
region, in case=1 and case=$s$, including and not the \bbr\ data.}
\em\hfill
\raisebox{2mm}{\bm{78mm}
\epsfig{file=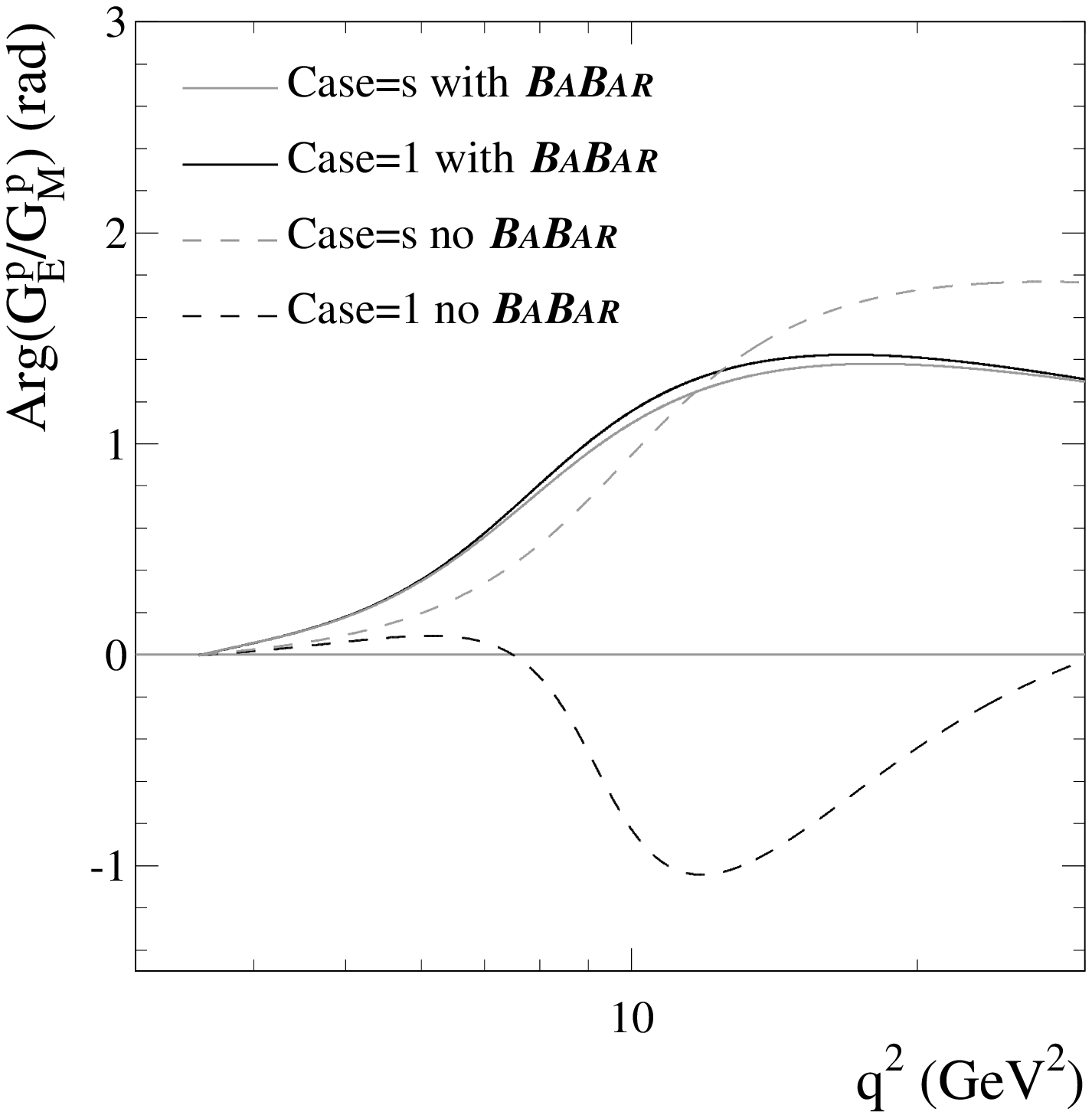,width=78mm}
\caption{\label{fig:rptl-ph} Prediction for the phase of the ratio $G_E^p/G_M^p$, in the time-like 
region, in case=1 and case=$s$, including and not the \bbr\ data.}
\em}
\ec\vspace{-3mm}
\efi
\section{Discussion}
\label{sec:discussion}
The Lomon-Gari-Krumpelman Model~\cite{lomon-list} was developed for and fitted to space-like 
EMFF data.  To enable the model to include the time-like region only the vector 
meson (of non-negligible width) propagators needed revision to appropriately 
represent a relativistic BW form at their pole in the time-like region.  
Two such forms are discussed above, case=1, the minimal alteration from the 
non-relativistic BW form, and case=$s$ derived from relativistic perturbation 
theory.  The resulting modification in the space-like region is minor and affected 
the fit there very little.
\\
With the new form of the vector meson propagators the simultaneous fit to the 
space-like EMFF and the time-like nucleon-pair production data was satisfactory 
as seen in Figs.~\ref{fig:gmp}-\ref{fig:geffn} and by the $\chi^2$ values of Table~\ref{tab:best-parms}. 
\\
The $\chi^2$ contributions from each space-like EMFF differ little between case=1 
and case=$s$ and are approximately the same as in the space-like only fit of Ref.~\cite{SL-data}. 
\\
However the fit in the time-like region, as measured by $\chi^2$, is qualitatively 
poorer when the \bbr\ data~\cite{babar} are included ($\chi^2$/d.o.f.=2.5) than when that 
set of data is omitted ($\chi^2$/d.o.f.=0.5 for case=1, and is 1.0 for case=$s$).  
As the quality of the fit is poorer when the \bbr\ data are included it may indicate 
an inadequacy in the model.  However the energy of the nucleon pairs produced in 
the \bbr\ experiment, unlike that of the exclusive pair production~\cite{fenice}-\cite{e835}, depends 
on the assumption that the observed photon is from electron or positron emission 
and is not accompanied by a significant amount of other radiation.  The resultant 
theoretical error is not fully known although relevant calculations have been made~\cite{egle}. 
The angular distributions may be sensitive to these radiation effects affecting the values 
of the $|G_E^p/G_M^p|$ ratio whose data are displayed in 
Fig.~\ref{fig:rptl} together with our prediction.
\\
Figures~\ref{fig:gmp}-\ref{fig:rn} are extended to higher momentum-transfers than the present data to show 
how the four different fits may be discriminated by new data.   Figure~\ref{fig:rp} for $R_p$  
indicates that at the higher momentum-transfers extended data may discriminate the 
smaller case=$s$ no \bbr\  prediction from the larger case=1 and case=$s$ with-\bbr\ predictions 
and from the still larger case=1 no-\bbr\ prediction.  Figure~\ref{fig:rn} for $R_n$ shows that at 
high momentum-transfer the case=$s$ predictions are higher than those for case=1.
\\
Figure~\ref{fig:geffn} is extended in energy for the same reason.  It clearly shows that at higher 
energy case=1 no-\bbr\  may be discriminated from the other three fits by 
moderately precise data.
\\
An extension of Fig.~\ref{fig:geffp} would only show the production of proton pairs 
remaining very close to zero.  However in the range of energy already covered it is evident that 
the case=$s$  no-\bbr\ result is difficult to reconcile with the \bbr\ data for 
$s=5-7$ GeV$^2$.  However for $s=7.5-8.5$ GeV$^2$ the no-\bbr\ fits are closer to 
the \bbr\ data than are the with-\bbr\ fits.
\\
Figures~\ref{fig:rptl} and~\ref{fig:rptl-ph} show that experiments in the time-like region for the ratio 
$|G_E^p/G_M^p|$ and the phase difference of $G_E^p$ and $G_M^p$ would be effective 
in discriminating between the models presented here and other models as well.
\clearpage
\renewcommand{\theequation}{A.\arabic{equation}}    
\setcounter{equation}{0}  %
\section*{Appendix A:\\
Dispersion Relations}
\label{app:dr}

Dispersion relations are based on the Cauchy theorem.
Consider a function $F(z)$, analytic in the whole $z$ complex plane 
with the discontinuity cut $(s_0,\infty)$.
If that function vanishes faster than $1/\ln|z|$ as $|z|$
diverges we can write the spectral representation
\be
F(z)=\frac{1}{\pi}\int_{s_0}^\infty\frac{\im[F(x)]dx}{x-z}\,.
\label{eq:drim}
\en
This is the so-called DR for the imaginary part where it is 
understood that the imaginary part is taken over the upper 
edge of the cut.\\
The extension to the case where there is a finite number of additional 
isolated poles  is quite natural. 
Indeed, considering a function with the set of $N$ 
poles $\{z_j\}$ ($j=1,\ldots,N$) of Sec.~\ref{subsec:regu}, under the same conditions
we obtain the spectral representation
\be
F(z)+2\pi i\sum_{j=1}^N{\rm Res}
\left[\frac{F(z')}{z'-z},z_j\right]=
\frac{1}{\pi}\int_{s_0}^\infty\frac{\im[F(x)]dx}{x-z}\,,
\label{eq:poles0}
\en
where Res$\big[g(z),z_0\big]$ stands for the residue of the
function $g(z)$ at $z=z_0$.
Furthermore, since we know the poles, we can use the more explicit
form
\be
F(z)=\frac{f(z)}{\prod_{j=1}^N(z-z_j)}\,,\no
\en
where $f(z)$ is the pole-free part of $F(z)$, but it has the same 
discontinuity cut. Using this form in the residue definition of
eq.~(\ref{eq:poles0}) and defining $\widetilde F(z)$ as the regularized 
version of $F(z)$, we have
\be
\widetilde F(z)=F(z)+\sum_{k=1}^N
\frac{f(z_k)}{\prod_{k=1,k\not=j}^N(z_k-z_j)}\frac{1}{z_k-z}
=
\frac{1}{\pi}\int_{s_0}^\infty\frac{\im[F(x)]dx}{x-z}\,
\no\label{eq:poles1}
\en
which is exactly the same expression as eq.~(\ref{eq:regu0}).
In other words, the DR procedure, using only the imaginary part 
of a generic function, which is suffering or not from the 
presence of unwanted poles, guaranties regularized analytic 
continuations, the poles, even if unknown, are automatically 
subtracted. 
\renewcommand{\theequation}{B.\arabic{equation}}    
\setcounter{equation}{0}  %
\section*{Appendix B:\\
A third case}%
\label{app:appb}%
We consider a regularized vector meson propagator~\cite{vmd-regu}
\be
D(s)=\frac{1}{M_0^2-s+\Pi(s)}\,,
\no\label{eq:d}
\en
where $M_0$ is the bare mass of the meson and $\Pi(s)$ is the scalar 
part of the tensor correlator. The imaginary part, due to the pion loop, 
can be obtained using the so-called Cutkosky rule~\cite{cutkosky} as
\be
\im\,\Pi(s)=-\gamma_{0}\, \sqrt{\frac{(s-\tilde s_0)^3}{s}}\,\theta(s-\tilde s_0)\,,
\hspace{10mm}\gamma_0=\frac{\Gamma_0 M^2}{\sqrt{(M^2-\tilde s_0)^3}}\,.
\label{eq:impi}
\en
The real part of $\Pi(s)$ represents the correction to the
bare mass $M_0$ in such a way that the dressed mass becomes
\be
M^2=M_0^2+\re\,\Pi(s)\,.
\no\label{eq:mass}
\en 
It follows that the propagator can be written in terms of $M^2$
and the only imaginary part of $\Pi(s)$
\be
D(s)=\frac{1}{M^2-s-i\,\im\,\Pi(s)}=\frac{1}{M^2-s-i\,\theta(s-\tilde s_0)\,
\gamma_0\,\sqrt{(s-\tilde s_0)^3/s}}\,.
\label{eq:original}
\en
Actually, only the imaginary part of this expression makes sense because of the 
Heaviside step function in the definition of eq.~(\ref{eq:impi}), nevertheless,
using DR, one can determine the complete propagator starting just
from its imaginary part. The propagator is expected to be real below the threshold $\tilde s_0$.
In particular, using eq.~(\ref{eq:drim}) for $t<\tilde s_0$, we have
\be
D(t)=\frac{1}{\pi}\int_{\tilde s_0}^\infty\frac{\im\, D(s)ds}{s-t}=
\frac{\gamma_0}{\pi}\int_{\tilde s_0}^\infty\frac{\sqrt{s(s-\tilde s_0)^3}\,ds}
{[s(M^2-s)^2+\gamma_0^2(s-\tilde s_0)^3](s-t)}\,,
\label{eq:dr}
\en 
while the real part over the time-like cut $(\tilde s_0,\infty)$, i.e. for $s>\tilde s_0$, is
\be
\re\, D(s)=\frac{1}{\pi}\Pr\!\!\int_{\tilde s_0}^\infty\frac{\im\, D(s')ds'}{s'\!-\!s}=
\frac{\gamma_0}{\pi}\Pr\!\!\int_{\tilde s_0}^\infty\frac{\sqrt{s'(s'\!-\!\tilde s_0)^3}\,ds'}
{[s'(M^2\!-\!s')^2\!+\!\gamma_0^2(s'\!-\!\tilde s_0)^3](s'\!-\!s)}\,.
\label{eq:dr-re}
\en 
In this case the ``natural'' space-like extension of the original form given in
eq.~(\ref{eq:original}) is no more possible, in fact such a form, when we forget
the Heaviside function in the denominator, develops a second cut which extends
over the whole space-like region. It follows that we can not write an expression 
like
\be
\mathbb{R}\ni \widetilde D(s<\tilde s_0)\not=\frac{1}{M^2-s-i\,\sqrt{(s-\tilde s_0)^3/s}}\,-
\underbrace{\sum_k\frac{R_{k}}{s-s_k}}_{\mbox{Physical poles}}\,,
\no\en 
where we get, in the space-like region, a regular and real propagator simply 
by subtracting the physical poles.
\\
The only possibility to go below threshold is to use the DR's of eq.~(\ref{eq:dr}) 
and~(\ref{eq:dr-re}). We compute explicitly the DR integrals
%
%Figures~\ref{fig:confsl} and~\ref{fig:conf} show comparisons among 
%the three descriptions in case of $\rho$ in the space-like and
%time-like regions respectively. On the left of each figure it
%is reported the modulus, on the right the relative difference 
%with respect to $BW_s$.\\
%
%
%A practical and handy analytic form for the propagator $D(s)$ 
%of eq.~(\ref{eq:original}) below threshold can be achieved by  
%integrating DR's of eq.~(\ref{eq:dr}) and~(\ref{eq:dr-re}) 
using the substitution
\be
x=\sqrt{1-\frac{\tilde s_0}{s}}\hspace{5mm}\Longrightarrow\hspace{5mm}
\left\{\begin{array}{l}
\displaystyle 
s=\frac{\tilde s_0}{1-x^2}\\
\\
\displaystyle 
ds=2\tilde s_0\frac{x\,dx}{(1-x^2)^2}\\
\\
\displaystyle 
s\in(\tilde s_0,\infty)\to x\in(0,1)\\
\end{array}\right.\,.
\no
\en
%
%%%The analytic form obtained for real and imaginary parts
%%%the regularized form for $D(s)$ is
The regularized form for $D(s)$ is
\be
%\widetilde{BW}_{\sqrt{s}}
\widetilde D(s)=
\!\left\{\!\!\begin{array}{ll}
\ds\frac{-1}{\pi\,s\,\gamma_0}
\sum_{i=0}^3\frac{\xi_i^3
\ln\left(\frac{\xi_i+1}{\xi_i-1}\right)}
{\prod_{k\not=i}^3(x_i^2-x_k^2)} &\hspace{4mm} s\le \tilde s_0\\
&\\
\ds-
%\frac{-1}{\pi\,s\,\gamma_0}
\frac{
%\left[
{\ds\sum_{i=1}^3}\frac{\xi_i^3
\ln\left(\frac{\xi_i+1}{\xi_i-1}\right)}
{\prod_{k\not=i}^3(x_i^2 - x_k^2)}+
\frac{x_0^3
\ln\left|\frac{\xi_0+1}{\xi_0-1}\right|}
{\prod_{k\not=0}^3(x_0^2 - x_k^2)}%\right]
}{\pi\,s\,\gamma_0}\!+\!
\ds \frac{i\,\gamma_0\sqrt{s(s\!-\!\tilde s_0)^3}}
{s(M^2\!-\!s)^2\!+\!\gamma_0^2(s\!-\!\tilde s_0)^3} &\hspace{4mm}s > s_0\\
\end{array}\right.\!\!\!\!.
\no\label{eq:3case}
\en
The four values $x_{i}^2$ ($i=0,1,2,3$), with $\xi_i\equiv\sqrt{x_i^2}$, 
are the roots of the  4$^{\rm th}$-degree 
polynomial in $x^2$, which represents the denominator of the integrands in
both DR's:
\be
\left\{\left[\left(1-x^2\right)M^2/\tilde s_0-1\right]^2/\gamma_0^2+x^6\right\}
\left(\tilde s_0/t-1+x^2\right)\,,
\label{eq:poly}
\en
in particular: $x_0^2=1-\tilde s_0/s$, is the only root that depends on $s$,
while the three $x_i$, with $i=1,2,3$, are the constant zeros of the first 
polynomial factor of two in eq.~(\ref{eq:poly}). The value at $s=0$ can be obtained 
as
\be
%%\widetilde{BW}_{\sqrt{s}}(0)&=&
\widetilde D(0)&=&
-\frac{1}{\pi\,s_0\,\gamma_0}
\sum_{i=1}^3\frac{x_i^3
\ln\left(\frac{x_i+1}{x_i-1}\right)}
{\mathop{\prod_{k\not=i}^3}_{k=1}(x_i^2-x_k^2)}\,.
\no
\en
Concerning the asymptotic behavior, when $s\to\pm\infty$, i.e.: $x_0^2\to 1$,
is
\be
%%\widetilde{BW}_{\sqrt{s}}(s)
\widetilde D(s)
\mathop{\sim}_{|s|\to\infty}
\frac{1}{\pi\gamma_0\prod_{i=1}^3(1-x_i^2)}\,
\frac{\ln|s|}{s}=
\frac{\gamma_0}{\pi(1+\gamma_0^2)}\,
\frac{\ln|s|}{s}\,,
\no\label{eq:asy3}
\en
where the last identity follows because the product
at denominator is just the $3^{\rm th}$ degree $x^2$-polynomial
of eq.~(\ref{eq:poly}) evaluated at $x^2=1$.
\\
A data fit was not made for this case because the resonance shape it 
produces is intermediate between the fitted case=1 and case=$s$.
\renewcommand{\theequation}{C.\arabic{equation}}    
\setcounter{equation}{0}  %
\section*{Appendix C:\\
The threshold behavior}%
\label{app:appc}%
The effective proton and neutron EMFF's extracted from the cross section 
data through the formula of eq.~(\ref{eq:geff-data}) have a quite steep enhancement
towards the threshold, i.e. when $q^2\to (2M_N)^2$. This is a consequence 
of the almost flat cross section measured in the near-threshold region:
$(2M_N)^2\le q^2\le (2\,\gev)^2$. Such a flat behavior is in contrast 
with the expectation in case of a smooth effective FF, which
gives, near threshold, a cross section proportional to the velocity of the 
outgoing nucleon $\sqrt{1-4M_N^2/q^2}$. Moreover, in the threshold region
the formula of eq.~(\ref{eq:geff-data}) has to be corrected to account for \NN\ 
finale state interaction. In particular, in the Born cross section formula,
in case of proton-antiproton, we have to consider the correction  due
to their electromagnetic attractive interaction~\cite{Baldini:2007qg}. 
Such a correction, having a very weak dependence on the fermion pair total 
spin, factorizes and, in case of pointlike fermions, corresponds to 
%an enhancement factor which represents 
the squared value of the Coulomb scattering wave function at the 
origin, it is also called Sommerfeld-Schwinger-Sakharov rescattering 
formula~\cite{coulomb}. Besides the Coulomb force also strong interaction
could be considered. Indeed, when final hadrons are produced almost at
rest they interact strongly with each other before getting outside the 
range of their mutual forces~\cite{meissner-threshold}.  Indeed there is evidence for near threshold quasi-bound \NN\  states with widths in the tens of MeV~\cite{Wycech}.  It follows that
EMFF values in this energy region are affected by different kinds of corrections
whose form and interplay are not well known. Hence
we decided to include in the present analysis only data above $q^2=4\,\gev^2$,
to avoid the threshold region.
\\
Figures~\ref{fig:respp} and~\ref{fig:resnn} show the residue data-over-fit for 
the proton and neutron effective FF's, respectively. They have been obtained
dividing the fit functions shown in Figs.~\ref{fig:geffp} and~\ref{fig:geffn} 
by the corresponding data on $|G_{\rm eff}^{p,n}|$. The threshold enhancement
of the proton data exceeds the fit by a factor of more then two and, in the 
neutron case, even within large errors, the factor is about three.
\bfi[h!]
\bc
\bm{78mm}
\epsfig{file=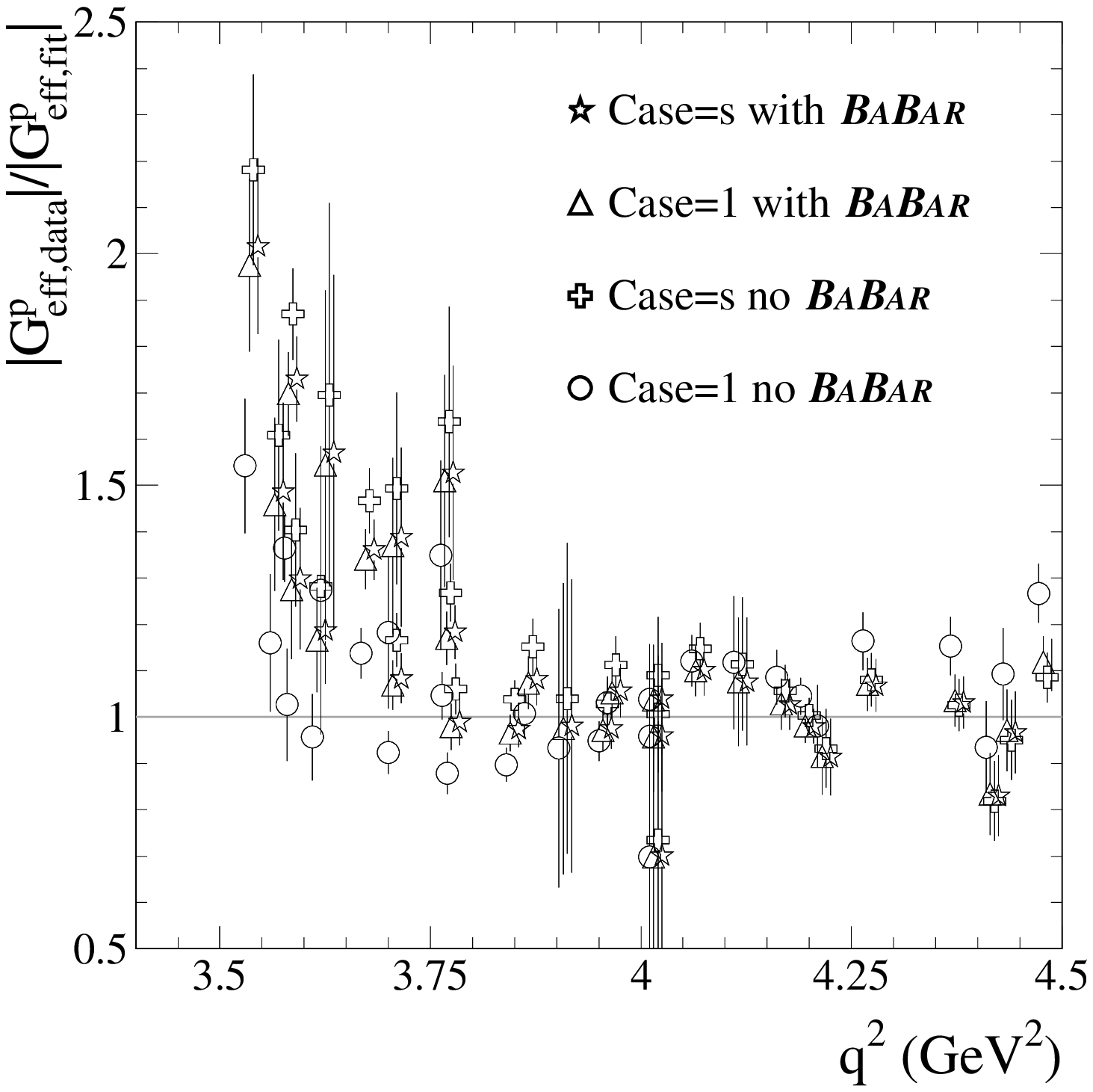,width=78mm}
\caption{\label{fig:respp}Residue for the proton effective FF:
$|G_{\rm eff,\,data}^p|/|G_{\rm eff,\, fit}^p|$, where $|G_{\rm eff,\, fit}^p|$
has been obtained considering only data with
$q^2\ge 4\,\gev^2$.}
\em\hfill
\bm{78mm}
\epsfig{file=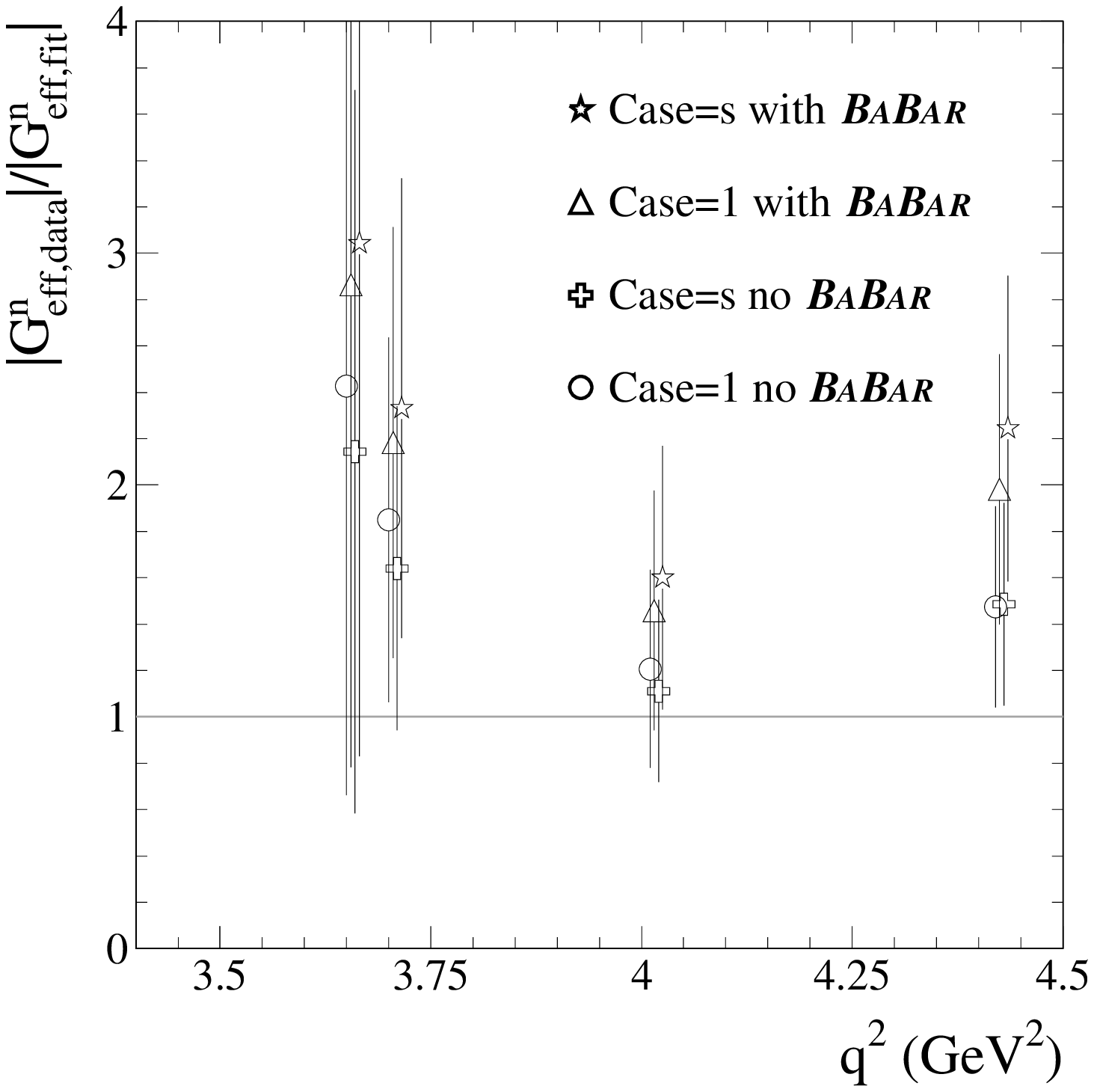,width=78mm}
\caption{\label{fig:resnn}\!Residue for the neutron effective FF:
$|G_{\rm eff,\,data}^n|/|G_{\rm eff,\, fit}^n|$, where 
$|G_{\rm eff,\, fit}^n|$ has been obtained considering only data 
with $q^2\ge 4\,\gev^2$.}
\em
\ec
\efi
%
%
%
%
%
%
%%%%%%%%%%%%%%%%%%%%%%%%%%%%%%%%%%%%%%%%%%%%%%%%%%%%%%%%%%%%%%%%%%%%%%%%
%

\end{document}